\documentclass[12pt,a4paper,notitlepage]{article}
\usepackage[a4paper, total={6.3in, 9.0in}]{geometry}
\usepackage{amsmath,amssymb,bm}
\usepackage{graphics}
\usepackage{epsfig}
\usepackage{graphicx}
\usepackage{color}
\usepackage{slashed}
\usepackage{amsbsy}
\usepackage{slashed}
\usepackage{xcolor}
\usepackage{framed}
\usepackage{float}
\definecolor{shadecolor}{RGB}{242,242,242}
\newcommand{\be}{\begin{equation} }
\newcommand{\ee}{\end{equation} }

\newcommand{\ket}[1]{\left| #1 \right\rangle}

\newcommand{\Lagr}{{\mathcal{L}}}

\newcommand{\pb}{\mathbf{p}}

\newcommand{\qb}{\mathbf{q}}
\newcommand{\bq}{\begin{eqnarray}}
\newcommand{\eq}{\end{eqnarray}}

\newcommand{\bsm}{\begin{small}}
\newcommand{\esm}{\end{small}}
\newcommand{\bs}{\begin{snugshade}\noindent\it}
\newcommand{\es}{\end{snugshade}}
\begin{document}
\begin{titlepage}
\title{\bf A Course in Amplitudes}\vskip 2mm
\date{}
\author{\Large Tomasz R.\ Taylor\thanks{\noindent Email: {\tt t.taylor@neu.edu}. On sabbatical leave from Department of Physics,  Northeastern University, Boston, MA 02115, United States of America. }\\[2mm]
Institute of Theoretical Physics, Faculty of Physics\\
 University of Warsaw, Poland}
\maketitle
\begin{abstract}\noindent
This a pedagogical introduction to  scattering amplitudes in gauge theories.
It proceeds from Dirac equation and Weyl fermions to
the two pivot points
of current developments: the recursion relations of Britto, Cachazo, Feng and Witten, and the unitarity cut method pioneered by Bern, Dixon, Dunbar and Kosower. In ten lectures, it covers the basic elements of on-shell methods.
\end{abstract}\thispagestyle{empty}\end{titlepage}
\renewcommand{\theequation}{\thesection.\arabic{equation}}
\tableofcontents
\section{Introduction}
This report is based on the lecture notes from the ``Amplitudes" course held at the Faculty of Physics of Warsaw University in Poland, in the winter semester of 2016/17. The course was sponsored by the Fulbright Scholar Program of the United States Department of State Bureau of Educational and Cultural Affairs. Its goal was to introduce upper level undergraduate students and beginner graduate students into the rapidly developing research area of scattering amplitudes.

Over the last thirty years many books and reviews appeared covering various aspects of this vast research area. Some of them, which are closely related to the present text, are listed at the end of this section. The goal of this review was to reach the two pivot points
of the current developments: the recursion relations of Britto, Cachazo, Feng and Witten, and the unitarity cut method pioneered by Bern, Dixon, Dunbar and Kosower. I believe that a student with a solid understanding of these two techniques is capable of moving to advanced topics, both on the ``formal'' side like the ideas of scattering equations and amplituhedron or on more ``applied'' side like the loop corrections in the standard model. As a starting point, I chose the classic textbook ``Introduction to Quantum Field Theory'' by Peskin and Schroeder, assuming that students have some basic understanding of perturbation theory - that is that they have seen before some Feynman diagrams and are familiar, here again to some extent, with QCD and  asymptotic freedom. Having determined these initial and final conditions, I was able to focus on the foundations of spinor helicity techniques, factorization techniques and the unitarity method, with the preference of  depth over breadth...

The main body of this report is divided into ten sections, each of them corresponding to an approximately 90-minute lecture. They include Exercises that fill some computational gaps. At the end of each section there is a short bibliography, but only of the articles and books most relevant to the material.
These are recommended readings so I tried to avoid distracting students with a long list of credits. {}For that reason I am asking in advance for forgiveness if I omitted some important references, but I will stick to my choice because it is a teaching material rather than a formal review.

I am grateful to students who attended my lectures, suggesting many improvements and corrections. I am particularly grateful to Fan Wei who double-checked many computations and  produced the figures.
I am grateful
to the United States Department of State Bureau of Educational and Cultural Affairs Fulbright
Scholar Program and to Polish-U.S. Fulbright Commission for a Fulbright Award
to Poland.
I acknowledge invaluable help during my sabbatical leave and encouragement for writing this report from Professor Zygmunt Lalak who hosted me at the Institute of Theoretical Physics of Warsaw University. This material is based in part upon work supported by the
National Science Foundation under Grant Number PHY-1620575.
Any opinions, findings, and conclusions or recommendations
expressed in this material are those of the author and do not necessarily
 reflect the views of the National Science Foundation.
 \renewcommand{\refname}{\large Recommended Books and Reviews}
  
\section{Units and Conventions}\subsection{Units}
The three basic units of Relativistic, Quantum, and Electromagnetic phenomena are:
\begin{center}
\renewcommand{\arraystretch}{2}
\begin{tabular}{c|c|c}
Relativistic&Quantum&Electromagnetic\\
\hline
$c=2.998\times 10^{8}\dfrac{m}{s}$&$\hbar=\dfrac{h}{2\pi}=1.055\times 10 ^{-34}J\cdot s$&$\varepsilon_{0}=\dfrac{1}{\mu_{0}c^{2}}=3.450\times 10^{26}\dfrac{e^{2}}{J\cdot m}$\\
\end{tabular}
\end{center}
where $e$ is the electron charge.
\noindent In this course, $c=\hbar=\varepsilon_{0}=1$. This then allows us to convert between space-time, mass-energy and charge, through $c$, $\hbar$, and $\varepsilon_{0}$. Note that in these units, the electron charge $e$ is dimensionless.
\bs {\bf Exercise 2.1}. We have the fine structure constant written as $\alpha=\dfrac{e^{2}}{4\pi}$ in our units of $c=\hbar=\varepsilon_{0}=1$ and we know that $\alpha$ is dimensionless. Now we wish to compute $\alpha$. To accomplish this the correct combination of $c$'s, $\hbar$'s and $\varepsilon_{0}$'s must be introduced. By inspection we find that $\alpha=\dfrac{e^{2}}{4\pi \varepsilon_{0} c \hbar}$ is dimensionless and we get $\alpha\approx 1/137$.\es
\subsection{Conventions}
Our conventions and notation follow the textbook ``Introduction to Quantum Field Theory'' by Peskin and Schroeder.
In particular, we are using the West Coast Minkowski metric $g_{\mu\nu}={\rm diag}(+,-,-,-)$. The three spatial components of the contravariant four-vector components with upper Lorentz indices coincide with the standard three-vector components, for example $x^\mu=(t,\mathbf{r})$ and $p^\mu=(E,\mathbf{p})$.
Under Lorentz transformations $x^\mu\to \Lambda^{\mu}_{~\nu}x^\nu$. For small transformations $\Lambda^{\mu}_{~\nu}\approx \delta^{\mu}_{\nu}+\omega^{\mu}_{~\nu}$. In order to identify $\omega^{\mu}_{~\nu}$, let us have a look at two examples:\\
1. Rotation in (12) plane
$$\left(\begin{array}{cc}
\cos\theta &-\sin\theta\\ \sin\theta &\cos\theta\end{array}\right)\quad \omega^{1}_{~2}=-\theta=-\omega^{2}_{~1}\Rightarrow\omega_{12}=-\omega_{21}=\theta. $$
2. Boost in (01) ``plane'' $$\left(\begin{array}{cc}
\cosh\beta &\sinh\beta\\ \sinh\beta &\cosh\beta\end{array}\right)\quad
\omega^{0}_{~1}=\beta=-\omega^{1}_{~0}\Rightarrow\omega_{01}=-\omega_{10}=\beta$$
These examples make it clear that $\omega_{\mu\nu}$ is antisymmetric.
Here $\beta$ is the rapidity: $\cosh\beta =1/\sqrt{1-v^2}\Rightarrow \tanh\beta=v$, so it should not be confused with the usual beta=v/c.

Two-component spinor notation is introduced in Sections 3 and 4.
\section {Dirac Equation: Spinors and $SL(2,C)$ Lorentz Symmetry}\subsection{Dirac Field}
\setcounter{equation}{0}

Spin 1/2 particles are quantum excitations of Dirac {\em spinor} fields. Dirac spinor has four components $\{\psi_a\}$, $a=1,2,3,4$, which can be collected in one column as
\be	\psi=
            \left(
            \begin{array}{lcr}
            \psi_1 \\
            \psi_2 \\
            \psi_3 \\
            \psi_4 \\
            \end{array}
            \right)\label{ds}
\ee
We are already familiar with the Dirac equation for spinor wavefunctions in relativistic quantum mechanics. Dirac equation is also the fundamental equation for the free Dirac field $\psi(x)$:
\begin{equation}
            (i\gamma^\mu\partial_\mu-m)\psi(x)=0 \label{deq1}
        \end{equation}
and will be used in the same way as the Klein-Gordon equation for scalar fields. Here, $\gamma^\mu$ are 4$\times$4 matrices  satisfying the so-called Clifford algebra of anticommutation relations:
\begin{equation}
        		\{\gamma^\mu,\gamma^\nu\}=2g^{\mu\nu} \label{gamma_commutation}
        \end{equation}
We also define the conjugate spinor
   \be	\bar\psi=\psi^\dagger\gamma^0=
            (
            \bar\psi_1 ,
            \bar\psi_2,
            \bar\psi_3 ,
            \bar \psi_4 )
           \ee

In Dirac theory, it is very handy to use the ``slashed'' symbol
\be
\slashed{v}=v_\mu\gamma^\mu~, \qquad \slashed{v}\slashed{v}=v^2\boldsymbol 1
\ee
where the second equation, with $\boldsymbol 1$ being a 4$\times$4 identity matrix, follows from Clifford algebra. In this way, Dirac equation is written as
\begin{equation}
            (i\slashed{\partial}-m)\psi=0 \label{de2}
        \end{equation}
It is clear that the solutions of Dirac equation must also satisfy the Klein-Gordon equation:
\begin{equation}
    0=(-i\slashed{\partial}-m)        [(i\slashed{\partial}-m)\psi]=(\partial^2+m^2)\psi \label{klein}
        \end{equation}

Under Lorentz transformations, Dirac spinors transform as
\be \psi(x)\to S_{\Lambda}\psi(\Lambda^{-1}x) \label{lol}\ee
where $S_\Lambda$ is a 4$\times$4 matrix acting on spinor indices:\footnote{The matrix $S_\Lambda$ (elements $(S_\Lambda)^a_b$) should not be confused with  $\Lambda$  (elements $\Lambda^{\mu}_\nu$) which is also
a 4$\times$4 matrix, but acting on \underline{vector} indices of spacetime coordinates and vector fields
[recall $A(x)\to\Lambda A(\Lambda^{-1}x)$].}
\be S_\Lambda=\exp\Big(-\frac{i}{2}\omega_{\mu\nu}\Sigma^{\mu\nu}\Big),\ee
\nopagebreak with
 \be \Sigma^{\mu\nu}=\frac{i}{4}[\gamma^\mu,\gamma^\nu]\ . \ee
In Eq.(\ref{lol}), $\omega_{\mu\nu}$ are the 6 parameters of Lorentz transformations. They represent 3 angles $\theta_i$ and 3 rapidity parameters $\beta_i$:
 \be \omega_{ij}=\epsilon_{ijk}\theta_k \qquad,\qquad \omega_{0i}=\beta_i\ .\label{angrap}\ee
\subsection{$SL(2,C)$ Lorentz Symmetry}
In order to discuss Lorentz transformations in more detail, it is convenient to use the Weyl representation for gamma matrices:
\be \label{gam}
            \gamma^\mu=
            \left(
                \begin{array}{lcr}
                    0                       &\sigma^\mu     \\
                    \overline{\sigma}^\mu   &0              \\
                \end{array}
            \right)
            ,~~~~
            \sigma^\mu=(\boldsymbol 1,\boldsymbol{\sigma}) ,~~~~  \overline{\sigma}^\mu=(\boldsymbol 1, -\boldsymbol{\sigma})
        \ee
where the vector $\boldsymbol\sigma$ consists of  3 Pauli matrices. We also define
\be \gamma^5=i\gamma^0\gamma^1\gamma^2\gamma^3=\left(
                \begin{array}{cc} -1&0\\0&1\end{array}\right)~,\qquad
                \{\gamma^5,\gamma^\mu\}=0
                \ee
It is easy to see:
\begin{itemize}
\item Squares: \be (\gamma^0)^2=(\gamma^5)^2=1~,~~~(\gamma^i)^2=-1\ee
\item Conjugation: \be (\gamma^\mu)^\dagger=\gamma^0\gamma^\mu\gamma^0~,~~~(\gamma^5)^\dagger=\gamma^5\ee
\end{itemize}

In Weyl representation, it is convenient to consider each two-component upper and lower block of a four-component Dirac spinor separately:
\be\psi=\left(\psi_L\atop\psi_R\right).\label{dlr}\ee
The two-component objects $\psi_L$ and $\psi_R$ are called left-handed and right-handed {\em Weyl spinors}. Formally, they can be extracted from the Dirac spinor by applying the projection operators:
\be
\left(\psi_L\atop 0\right)=P_L\psi~,~ \left(0\atop \psi_R\right)=P_R\psi~~ {\rm with}~~
P_{L}=\frac{1-\gamma^5}{2}~,~P_{R}=\frac{1+\gamma^5}{2}\ .
\ee
\bs
\noindent{\bf Exercise 3.1}  Show that under Lorentz transformations: rotations by $\boldsymbol\theta=(\theta_1,\theta_2,\theta_3)$ and boosts by $\boldsymbol\beta=
(\beta_1,\beta_2,\beta_3)$, Weyl spinors transform as
\be\psi_L\to \exp(\boldsymbol -z\cdot \frac{\boldsymbol{\sigma}}{2})\psi_L~,\quad \psi_R\to \exp(\boldsymbol{z^*}\cdot \frac{\boldsymbol\sigma}{2})\psi_R,\quad{\rm where}~\boldsymbol z=\boldsymbol \beta+i\boldsymbol \theta\ .\label{slc}\ee
\es
Since upper and lower components of the Dirac spinor do not mix under Lorentz transformations, {\em c.f}.\ Eq.(\ref{slc}), Dirac spinor is said to be in a {\em reducible\/} representation of the Lorentz group. Furthermore, acting on $\psi_L$, Lorentz transformations are represented by
2$\times$2 matrices
\be M(\boldsymbol z)=\exp(\boldsymbol -z\cdot \frac{\boldsymbol{\sigma}}{2})\ .\ee
\bs
{\bf Exercise  3.2} {\it Prove that $M(\boldsymbol z)$ has the form of a most general complex matrix with  determinant 1. Show that $M^{-1}=-\epsilon M^T\epsilon$, where $\epsilon=i\sigma_2$ is the 2$\times$2 antisymmetric symbol with $\epsilon^2=-\boldsymbol 1$.}
\es\noindent
We conclude that the Lorentz group $SO(1,3)$ is (locally) isomorphic to the special linear group $SL(2,C)$. Two-component left-handed spinors $\psi_L$ transform in the fundamental {\bf 2} representation of $SL(2,C)$: $\psi_L\to M\psi_L$. Regarding the right-handed $\psi_R$, it is more convenient to consider
$\epsilon\psi_R$. These spinors transform in the antifundamental $\boldsymbol{\bar{2} }$ representation: $\epsilon\psi_R\to M^*\epsilon\psi_R$, which can be demonstrated by using $\epsilon\boldsymbol\sigma\epsilon=\boldsymbol\sigma^*$. To summarize, a Dirac spinor can be written as
\be\psi=\left(\chi_\alpha\atop\bar\eta^{\dot\alpha}\right)~,\quad
\bar\eta^{\dot\alpha}\equiv\epsilon^{\dot\alpha\dot\beta}\bar\eta_{\dot\beta}
\label{dtc}\ee
where the undotted and dotted subscript indices $(\alpha,\dot\beta=1,2)$ distinguish between {\bf 2} and $\boldsymbol{\bar{2}}$  representations of Lorentz $SL(2,C)$. Under Lorentz transformations,
\be \chi_{\alpha}\to (M\chi)_{\alpha}\ ,\qquad \bar\eta^{\dot\alpha}\to (M^{\dagger-1}\bar\eta)^{\dot\alpha}.\label{ltrs}\ee
We will be also often representing four-vectors as $2{\times}2$ matrices:
 \be v_{\alpha\dot\beta}\equiv v_\mu(\sigma^\mu)_{\alpha\dot\beta}\ ,\qquad\bar v^{\dot\alpha\beta}\equiv v_\mu(\bar\sigma^\mu)^{\dot\alpha\beta}\ .\ee
 More explicitly,
$$v=\left(\begin{array}{cc}
v_0+v_3 &v_1-iv_2\\ v_1+iv_2 &v_0-v_3\end{array}\right)\quad
\bar v=\left(\begin{array}{cc}
v^0+v^3 &v^1-iv^2\\ v^1+iv^2 &v^0-v^3\end{array}\right)\ .  $$
Note that $\det v=\det\bar v=v^2$. In this notation, the Dirac equation (\ref{deq1}) reads:
\be i\partial\bar\eta-m\chi ~=~ 0 ~=~ i\bar{\partial}\chi-m\bar\eta=0\ .\label{deq}\ee
\bs
{\bf Exercise 3.3} {\it
 Show that under Lorentz transformations $v^\mu\to \Lambda^{\mu}_{~\nu}v^\nu$}
\be   v\to M\,  v\,  M^{\dagger}\ ,\qquad \bar v\to M^{\dagger -1}\bar v\, M^{-1}\ . \label{vltran}   \ee
{\bf Exercise 3.4} {\it Let $v^\mu$ and $w^\mu$ be  Lorentz vectors. Show that}
 \be {\rm Tr}(v\bar w)=2v^\mu w_\mu\equiv 2vw\ .\ee\es
\noindent As a corollary, we can establish Lorentz symmetry of Dirac theory.
\bs {\bf Exercise 3.5} Assume that $\psi(x)=(\chi,\bar\eta)(x)$ satisfies the Dirac equation (\ref{deq}).
Show that it is
 also satisfied by the Lorentz-transformed spinor $\psi'(x')=(\chi',\bar\eta')(x')$ with
\be \chi'(x')=M\chi(\Lambda^{-1}x)\, \quad
\bar\eta'(x')=M^{\dagger-1}\bar\eta(\Lambda^{-1}x)\ee\es
 \renewcommand{\refname}{\large Recommended Reading for Section 3}
  
\section{Solutions of Dirac Equation and Chiral Fermions}\setcounter{equation}{0}
\subsection{Non-zero Mass}

To find the solutions of Dirac equation (\ref{deq1}), it is convenient to rewrite it in terms of $\psi_L$ and $\psi_R$:
\be\left(\begin{array}{cc} -m & i(\partial_0+\boldsymbol\sigma\cdot\boldsymbol\nabla)\\ i(\partial_0-\boldsymbol\sigma\cdot\boldsymbol\nabla)& -m\end{array}\right)\left(\begin{array}{c}\psi_L\\ \psi_R\end{array}\right)=0.\ee
Since the solutions must also satisfy the Klein-Gordon equation, we should be able to construct them as a superposition of ``plane waves"
\be\psi_p(x)=u(\pb)e^{-ipx}+v(\pb)e^{ipx}\ .\label{psisol}\ee
{}For the ``positive energy'' solution $u(\pb)$, the above equation translates to
\be mu_R(\pb)=(E_p+\pb\cdot\boldsymbol \sigma)u_L (\pb)\label{posol}\ee
where $E_p=\sqrt{\pb^2+m^2}$,
and the second set, with $R\leftrightarrow L$ and $\pb\to -\pb$, which is redundant because $p^2=m^2$. It is completely sufficient to solve it in the rest frame, where $\pb=0$ and $E=m$,
 because a solution with arbitrary momentum can be obtained by appropriate boost and/or rotation. There are two solutions:
  \be u^1_L(0)=\sqrt{m}\left(   \begin{matrix} 1\\ 0\\ \end{matrix}  \right)=u^1_R(0)\ ,\qquad
  u^2_L(0)=\sqrt{m}\left(   \begin{matrix} 0\\ 1\\ \end{matrix}  \right)=u^2_R(0)\ ,\label{usol}\ee
where we used the normalization adopted in most of textbooks. Similar ``negative energy" solutions are
\be v^1_L(0)=\sqrt{m}\left(   \begin{matrix} 0\\ 1\\ \end{matrix}  \right)=-v^1_R(0)\ ,\qquad
  v^2_L(0)=\sqrt{m}\left(   \begin{matrix} -1\\ 0\\ \end{matrix}  \right)=-v^2_R(0)\ ,\label{vsol}\ee
Note that $u^{1,2}(0)$ are spin``up'' and ``down'' eigenstates ($s_z=\pm 1/2$), respectively of the spin operator $\Sigma^{12}=S_3$. For the negative energy solutions, the spin assignment is flipped because of their antiparticle interpretation, but there is no need to discuss it at this point.
\bs {\bf Exercise 4.1} Prove the following completeness relations:
 \be  \sum_{s=1,2} u_a^s(p)\bar{u}_b^s(p)=(\slashed{p}+m\boldsymbol 1)_{ab}~~~~~ \makebox{($a,b=1,2,3,4$ are spinor indices)}\label{co1}\ee
 \be  \sum_{s=1,2} v_a^s(p)\bar{v}_b^s(p)=(\slashed{p}-m\boldsymbol 1)_{ab}~~~~~ \makebox{($a,b=1,2,3,4$ are spinor indices)}\label{co2}\ee
 \es
The free Dirac field operators have the form
\begin{eqnarray}\psi(x) &=& \int \frac{d^3\pb}{(2\pi)^3}\frac{1}{\sqrt{2E_p}}\sum_{s=1,2}[a_{\pb}^su^s(\pb)e^{-ipx}+
b_{\pb}^{s\dagger}v^s(\pb)e^{ipx}]\label{freed}\\
\bar\psi(x) &=& \int \frac{d^3\pb}{(2\pi)^3}\frac{1}{\sqrt{2E_p}}\sum_{s=1,2}[
b_{\pb}^{s}\bar v^s(\pb)e^{-ipx}+a_{\pb}^{s\dagger}\bar u^s(\pb)e^{ipx}]\label{freedc}\end{eqnarray}
    with $E_p=\sqrt{\pb^2+m^2}$, so that $p^2=m^2$.
The basic difference between fermions (half-integer spin) and bosons (integer spin) appears at the very outset of the quantization: instead of commutation relations, we impose {\em anti}commutation relations for the creation and annihilation operators:
\be \{ a^r_\pb, a^{s\dagger}_\qb\}=\{ b^r_\pb, b^{s\dagger}_\qb\}~=~(2\pi)^3\delta^3(\pb-\qb)\,\delta^{rs}\label{com1}\ee
\be\{ a^r_\pb, a^{s}_\qb\}=\{ a^{r\dagger}_\pb, a^{s\dagger}_\qb\}= \{ b^r_\pb, b^{s}_\qb\}~=~\{ b^{r\dagger}_\pb, b^{s\dagger}_\qb\}~ =~0\label{com2}\ee

 The vacuum state is annihilated by all annihilation operators,
 \be a^s_\pb|0\rangle= b^s_\pb|0\rangle=0\ee
The lowest non-zero energy states are obtained by acting on the vacuum with a single creation operator. There are two types of particles created in this way, by $a^\dagger$ and $b^\dagger$, and we call them plus (+) (or electron) and minus ($-$) (or positron) states, respectively:
 \be |p,s,+\rangle=\sqrt{2E_p}\,a_p^{s\dagger}|0\rangle~,~~\qquad|p,s,-\rangle=\sqrt{2E_p}\,
 b_p^{s\dagger}|0\rangle\label{states}\ee
 The exercise below elaborates on the connection between Fermi-Dirac statistics, Pauli exclusion principle, and the anticommuting nature of Dirac fields.
 \bs
{\bf Exercise 4.2} {\it Consider a two-electron state
$\ket{e(p_1,s_1),e(p_2,s_2)}$. Show that it obeys Fermi-Dirac statistics:
$\ket{e(p_2,s_2),e(p_1,s_1)}=-\ket{e(p_1,s_1),e(p_2,s_2)}$}\es

In order to make a connection between left- and right-handed components of Dirac fields and the spin states of the particles they create (or annihilate), consider the wavefunction $u^1(\pb)$ describing a $s_z=+1/2$ particle in the rest frame. We are interested in the ultra-relativistic limit of this wave-function, when the particle moves with very large $p_z\gg m$ along the $z$-axis, with the spin pointing in its direction of motion. It can be obtained by boosting with $\beta_3\equiv\beta\to\infty$:
$$u^1_L(\pb)=e^{-\beta/2}u^1_L(0)\ ,\quad u^1_R(\pb)=e^{\beta/2}u^1_R(0)\ ,$$
therefore
\be u^1(\pb)=e^{\beta/2}P_R u^1(0)+{\cal O}(e^{-\beta/2}) ~\ \stackrel{\scriptscriptstyle E\gg m}{\longrightarrow} ~P_Ru^1(\pb)   .\ee
We see that the wave-function of an ultra-relativistic spin 1/2 particle with the spin pointing in the direction of motion, that is of a particle with a positive (right-handed) ``helicity'' $h=\hat\pb\cdot\mathbf{s}=1/2$, where $\hat\pb$ is the unit vector pointing in the direction of $\pb$, is indeed described by a purely right-handed spinor. Similarly, an ultra-relativistic  particle with
$h=-1/2$
is described by a left-handed wavefunction.
\bs{\bf Exercise 4.3} {\em Show that}
$\langle 0|\psi(x)|p,s,+\rangle =u^s(\pb)e^{-ipx}\label{ultra} .$\es
\noindent From the above ``reduction formula'', it follows that at very high energies
\be \langle 0|\psi(x)|p,h= {+}1/2,+\rangle\ ~\stackrel{\scriptscriptstyle E\gg m}{\longrightarrow}~\langle 0|P_{R}\psi(x)|p,h={+}1/2 ,+\rangle \ee
and a similar limit with $h={-}1/2$ and $P_L$.
Hence in the ultra-relativistic limit, left- and right-handed fermions are annihilated by the ``chiral''\footnote{The term ``chiral'' is derived from the Greek word $\chi\varepsilon\iota\rho$ (cheir) for ``hand."} fields $\psi_L(x)$ and $\psi_R(x)$, respectively.
\subsection{Zero Mass}
The case of massless fermions is very different from the massive case. Now the ``positive" and ``negative" energy waves in
\be\psi_p(x)=u(\pb)e^{-ipx}+v(\pb)e^{ipx}\qquad (p^2=0)\ .\label{psisoll}\ee
satisfy the same equation, i.e.\ $u(\pb)\sim v(\pb)$. Furthermore, this equation splits into two decoupled equations for the left- and right-handed spinor components, see Eq.(\ref{posol}):
\be (E_p+\pb\cdot\boldsymbol \sigma)u_L (\pb)~=~0~=~ (E_p-\pb\cdot\boldsymbol \sigma)u_R  \label{posoll}\ee
This means that one does not necessarily need a four-component Dirac spinor to describe a massless fermion.
Instead, one can consider two-component Weyl fields equivalent to four-component Dirac spinors with the upper or lower components constrained to be zero:
\be{\rm left\makebox{-}handed}~~ \chi_{W}=\left(\chi_\alpha\atop 0\right)~~ {\rm or~right\makebox{-}handed}~~ \bar\eta_{W}=\left( 0\atop
\bar\eta^{\dot \alpha}\right)\ .
\label{dtwc}\ee
Now the Dirac equation (\ref{deq}) reads
\be i\bar{\partial}\chi ~=~ 0 ~=~ i\partial\bar\eta\ .\label{deqw}\ee
We will see that such pure left- or right-handed fields describe fermions with two degrees of freedom.

As an example, consider a chiral theory of a free left-handed Weyl particle described by
\be\chi_\alpha(x) = \int \frac{d^3\pb}{(2\pi)^3}\frac{1}{\sqrt{2E_p}}\,\lambda_{\alpha}({\pb})\big(a_{\pb}e^{-ipx}+
b_{\pb}^{\dagger}e^{ipx}\big)\, \label{freew}\ee
where $E_p=|{\pb}|$. Now Eqs.(\ref{deqw}) read
\be\bar p^{\dot\alpha\beta}\lambda_{\beta}({\pb})=0~~~{\rm with}~~ \bar p^{\dot\alpha\beta}
=\left(\begin{array}{lcr}    E_p+p^3 & p^1-ip^2\\ p^1+ip^2& E_p-p^3 \end{array}\right).
\label{debar}\ee
Here $\det\bar p=p^2=0$. The above equation transforms covariantly under Lorentz transformations:
$\lambda\to M\lambda$ and $\bar p\to M^{\dagger -1}\bar p\,  M^{-1}$, see Eq.(\ref{vltran}).
We can use this covariance in order to solve Eq.(\ref{freew}) in a frame in which the particle moves along the positive $z$ direction, with some reference energy $E_0$, so that $E_p-p^3=p^1=p^2=0$ and $E_p+p^3=2E_0$. In this frame, Eq.(\ref{debar}) is trivially solved by
\be\lambda_{\alpha}(E_0)=\left( {0\atop \sqrt{2E_0}}\right).\ee
The general solution $\lambda_{\alpha}(\pb)$ can be obtained by rotating the frame and adjusting the energy by appropriate boost. $\lambda_{\alpha}(\pb)$ is usually called the left-handed  {\bf momentum spinor} while
$\bar\lambda^{\dot\alpha}(\pb)$ is the right-handed {\bf momentum spinor}.
\bs {\bf Exercise 4.4} {\em Show that }
\be\lambda_{\alpha}(\pb)\bar\lambda_{\dot\beta}(\pb)=p_{\alpha\dot\beta}\quad{\rm and}\quad
p_{\alpha\dot\beta}\bar\lambda^{\dot\beta}(\pb)=0\ .\label{pspin}\ee\es
Transforming the reference four-momentum to an arbitrary light-like vector specified by three parameters employs rotations by two angles and one boost. This means that for each four-momentum, there exists  a three-parameter $(6-3=3)$ subgroup of the Lorentz group that leaves it unchanged. It is called the little (Wigner) group. For a light-like momentum pointing in the $z$-direction, it is the rotation about the $z$-axis generated by $\Sigma^{12}=S_3$ and two hybrid transformations generated by
\be P_1=\Sigma^{01}-\Sigma^{31}~,\quad P_2=\Sigma^{02}-\Sigma^{32}\ .\ee
Indeed, by using Eq.(\ref{lol}) one finds
\be e^{i\delta_1P_1}\lambda(E_0)=e^{i\delta_2P_2}\lambda(E_0)=\lambda(E_0)\ ,\quad e^{i\theta S_3}\lambda(E_0)=e^{-i\theta/2}\lambda(E_0)\ .\ee
While $P_1$ and $P_2$ transformations leave the momentum spinors invariant, rotations by an angle $\theta$ about the momentum axis introduce phase factors which are determined by the helicity ($h=\hat\pb\cdot\mathbf{s}$): $e^{i\theta h}=
e^{\mp i\theta /2}$ for left- and right-handed spinors respectively.\footnote{The algebra of $P_1, P_2, S_3$ little group generators is isomorphic to $E_2$, the euclidean symmetry group of a two dimensional plane.} These factors cancel in the momentum vector, see Eq.(\ref{pspin}).

Going back to Eq.(\ref{freew}), we conclude that Weyl spinor fields describe massless fermions with a definite helicity $h=\hat\pb\cdot\mathbf{s}=\pm 1/2$, plus or minus one-half. The second degree of freedom describes antiparticles which can be shown to carry opposite helicities. In the framework of Standard Model, neutrinos are associated to left-handed Weyl fields, although the recent discovery of neutrino masses suggests that this description may need some modifications.
\subsection{More on Lorentz Symmetry and Invariants}
We want to learn how to construct Lorentz-invariant expressions involving chiral spinors. From Eq(\ref{ltrs}), we know that under Lorentz transformations
\be \chi_{\alpha}\to (M\chi)_{\alpha}\ ,\qquad \bar\eta^{\dot\alpha}\to (M^{\dagger-1}\bar\eta)^{\dot\alpha}.\ee
We will be often using the $\epsilon$ matrix (antisymmetric symbol) to raise and lower the spinor indices:
\be \chi^{\alpha} =\epsilon^{\alpha\beta}\chi_{\beta}\ ,\qquad \bar\eta_{\dot\alpha}=  \epsilon_{\dot\alpha\dot\beta}\bar\eta^{\dot\beta}\ee
with
\be\epsilon^{12}=\epsilon^{\dot{1}\dot{2}}=\epsilon_{21}=\epsilon_{\dot 2\dot 1}=+1\ ,\qquad
\epsilon^{21}=\epsilon^{\dot{2}\dot{1}}=\epsilon_{12}=\epsilon_{\dot 1\dot 2}=-1\ .
\ee
\bs
{\bf Exercise 4.5} Show that under Lorentz transformations
\be \chi^{\alpha}\to (\chi M^{-1})^{\alpha}\ ,\qquad \bar\eta_{\dot\alpha}\to (\bar\eta M^{\dagger})_{\dot\alpha}.\ee
\es
We can construct Lorentz invariants by contracting lower and upper indices. It is not possible to construct a Lorentz invariant from one spinor because obviously  $\chi^{\alpha}\chi_{\alpha}=\epsilon^{\alpha\beta}\chi_{\alpha}\chi_{\beta}=0$ etc.\ (these spinors are ordinary complex-valued functions). But if we have two spinors, the following product is Lorentz invariant:
\be \lambda_1^{\alpha}\lambda_{2\alpha }\equiv\lambda_1\lambda_2 \to \lambda_1M^{-1}M\lambda_2=\lambda_1\lambda_2
\ee
This product is called the ``angle" product of left-handed spinors
\be \langle 12\rangle\equiv \lambda_1\lambda_2 \ee
We can also construct a similar ``square" product of right-handed spinors:
\be [ 12]\equiv \bar\lambda_{1\dot\alpha}\bar\lambda_2^{\dot\alpha}\equiv  \bar\lambda_1\bar\lambda_2 \ee
These products are antisymmetric:
\be  \langle 21\rangle =- \langle 12\rangle\ ,\qquad  [21]=-[12]\ .\ee
Note that under complex conjugation,
\be\langle 12\rangle^*=[21]\ , \qquad  \langle 12\rangle[21]=|\langle 12\rangle|^2  \ee
\bs {\bf Exercise 4.6} Show that for two momentum spinors corresponding to $p_1$ and $p_2$, \be \langle 12\rangle [21]={\rm Tr}(\bar{p}_1p_2)=2p_1p_2\ .\ee
{\bf Exercise 4.7} Consider three momentum spinors $\lambda_{1,2,3}$. Since these are two-component vectors, each of them can be expressed as a linear combination of the other two:
$$\lambda_1=a\lambda_2+b\lambda_3$$
Determine $a$ and $b$.\es
\noindent As a corollary, we obtain very important ``{\bf Schouten's identity}":
\be\langle 14\rangle\langle 23\rangle=\langle 13\rangle\langle 24\rangle-\langle 12\rangle\langle 34\rangle\ .\label{schout}\ee
\renewcommand{\refname}{\large Recommended Reading for Section 4}
  
\section{Gauge Vector Bosons}\setcounter{equation}{0}
\subsection{Electromagnetic Fields}
Until this point, we discussed fermions, in particular Weyl fields
which represent two degrees of freedom of a left-handed $(h=-1/2)$ particle and right-handed  $(h=+1/2)$ antiparticle. We also discussed Dirac fields which are suitable for describing electrons and quarks. Now we turn to photons which in the framework of Quantum Electrodynamics (QED) appear as quanta of electromagnetic radiation.

A plane electromagnetic wave consists of electric and magnetic fields, $\mathbf{E}$ perpendicular to $\mathbf{B}$, oscillating in the plane transverse to the direction of the wave-vector, so that $\mathbf{E} \times \mathbf{B}$ points in the direction of propagation. The magnitude $|\mathbf{B}|=|\mathbf{E}|$, therefore the number of degrees of freedom is equal to the number of the degrees of freedom of the electric field which can oscillate in two transverse directions. In classical electrodynamics, these two degrees of freedom correspond to two possible polarizations. For a plane wave moving in $z$ direction we can have two linear polarizations in $x$ or $y$ directions, or equivalently, two circular: left- and right-handed polarizations. This means that in QED, the photon must have two degrees of freedom. One would think that the most straightforward way of constructing QED is to quantize the electric field.
Unfortunately this is not the case. One of the reasons is that since e-m waves move at the speed of light, the theory must be relativistic, but  $\mathbf{E}$ and $\mathbf{B}$ do not transform in a simple way under Lorentz transformations.

In relativistic electrodynamics, electric and magnetic fields are incorporated into one relativistically covariant tensor, the antisymmetric field-strength tensor:
\begin{equation}
F_{\mu \nu}\equiv \left(\begin{array}{rrrr}
0   &E_1&E_2 &E_3\\
-E_1&0     & -B_3& B_2 \\
-E_2&B_3 &0      &-B_1\\
-E_3&-B_2&B_1  & 0
\end{array}\right).
\label{ftensor}
\end{equation}
so that $E_i=F_{0i}$ and $B_i=-\frac{1}{2}\epsilon_{ijk}F^{jk}$.
Under Lorentz transformations,
\begin{equation}
F_{\mu \nu} \rightarrow \varLambda_{\mu}^{\rho} \varLambda_{\nu}^{\lambda} F_{\rho \lambda}.
\label{ftransformation}
\end{equation}
In the absence of external sources, which is completely sufficient for studying e-m waves propagating in the vacuum, Maxwell's equations read
\be \partial_{[\rho}F_{\mu\nu]}=0~~\mathbf{(M1)}\qquad \partial_{\mu}F^{\mu\nu}=0~~\mathbf{(M2)}\nonumber\label{meq}\ .\ee
In order to reduce the number of degrees of freedom in a way consistent with Maxwell's equations, one introduces the relativistic four vector
\be A^{\mu}=(A^0, \mathbf{A})\ee
which combines the standard electric potential and magnetic vector potential in one four-vector potential. In this formalism, the field-strength tensor is constructed as
\begin{equation}
F_{\mu \nu} =\partial_{\mu} A_{\nu}-\partial_{\nu} A_{\mu},
\label{elec}
\end{equation}
which automatically takes care of $\mathbf{(M1)}$. In order to obtain non-trivial solutions of $\mathbf{(M2)}$, we can use the so-called ``radiation gauge'' and set $A^0=0$. We are left with
\be \Box A^i=0~~~(\mathbf{KG})\ ,\qquad\partial_iA^i=0~~~(\mathbf{TT})\ .\label{dal}\ee
From here, it is very easy to obtain wave solutions. For example, the ansatz $\mathbf{A}=[A_x(t-z), A_y(t-z), 0]$ yields a wave propagating in the $z$ direction at the speed of light. These solutions, as well as Eqs.(\ref{dal}) show explicitly that the two electromagnetic degrees of freedom can be encoded in  two transverse $(\partial_iA^i=0)$ components of the vector potential whose space-time dependence is governed by the relativistic Klein-Gordon wave equation $(\mathbf{KG})$ with zero mass. Describing e-m fields in terms of four-potentials is a step in the right direction, but it is still not satisfactory because our discussion used a non-relativistic radiation gauge. In order to keep explicit Lorentz covariance we need to consider the full vector four-potential $A_\mu$ which contains two redundant (unphysical) degrees of freedom. One of these redundant degrees of freedom is easy to identify:
 {\bf gauge transformations}  $A_\mu\to A_\mu+\partial_{\mu}\Lambda$, where $\Lambda$ is an arbitrary function, do not change the physical $\mathbf{E}$ and $\mathbf{B}$ fields, as it is clear from Eq.(\ref{elec}). The second one is more tricky and is usually expressed as ``the scalar potential is not a dynamical field''. In any case, it is clear that the quantization of four-potentials, which are usually called {\bf gauge fields}, is a rather complex problem, so it is not surprising that it took many years to develop QED.
\subsection{Polarization Vectors}
 We start from  the plane wave
\be A^{\mu}_p(x)=\sum_{s=\mp}\big[\epsilon_s^{\mu}(p)e^{-ipx}a_{\mathbf{p}}^s+\epsilon_{s}^{*\mu}(p)e^{ipx}
a_{\mathbf{p}}^{\dag s}\big]\label{ggf}\ee
where $a_p$ are arbitrary (complex) amplitudes that will be later upgraded to annihilation operators. These waves satisfy $(\mathbf{KG})$ of Eqs.(\ref{dal}) provided that $p^2=0$. In the radiation gauge $\epsilon_{\pm}^{0}(p)=0$ and $(\mathbf{TT})$ is equivalent to $\mathbf{p}\cdot\boldsymbol{\epsilon}=0$. $\epsilon_{\pm}^{\mu}(p)$ are its two independent solutions describing left- and right-handed (circular polarizations) waves.  First consider such a plane wave moving in the $z$ direction with the momentum $p_0=E_0(1,0,0,1)$. In order to satisfy Eqs.(\ref{dal}), it must be a superposition of the waves defined by the polarization vectors
\be \epsilon_{\pm}^{\mu}(p_0)=  \frac{1}{\sqrt{2}} \left(
            \begin{array}{ccc}
            0 \\ 1 \\   \pm i \\ 0 \\    \end{array} \right).\label{eps}\ee
Note that these are complex, light-like vectors, $\epsilon_{\pm}^2=0$. They satisfy $p_{0\mu}\epsilon_{\pm}^{\mu}(p_0)=0$ and $r_{0\mu}\epsilon_{\pm}^{\mu}(p_0)=0$, where $r_0=(1,0,0,{-}1)$. \bs
{\bf Exercise 5.1} Prove that the matrix $\epsilon_{\alpha\dot\beta}=\epsilon_\mu(\sigma^\mu)_{\alpha\dot\beta}$ associated to any light-like (complex) vector $\epsilon_\mu$  can be written as a product of left-and right-handed spinors:
$$\epsilon_{\alpha\dot\beta}=\omega_{\alpha}\bar\rho_{\dot\beta}\ .$$ For a real vector, $\rho=\omega$.\\
{\bf Exercise 5.2} Find a spinor $r_{0\alpha}$ such that $r_{0\alpha}\bar r_{0\dot\beta}=r_{0\alpha\dot\beta}$ and
\be
\epsilon_{-}^{\mu}(p_0)=  \frac{1}{\sqrt{2}}\frac{\bar r_{0\dot\alpha}(\bar\sigma^{\mu})^{\dot\alpha\beta}\lambda_{0\beta}}{[ \lambda_0r_{0}]} ~,\qquad \epsilon_{+}^{\mu}(p_0)=\frac{1}{\sqrt{2}}\frac{r_{0}^{\alpha}(\sigma^{\mu})_{\alpha\dot\beta}\bar \lambda^{\dot\beta}_0}{\langle r_{0} \lambda_0\rangle}\label{ezero}
\ee
\es
In order to obtain polarization vectors for arbitrary momentum $p$, we use Lorentz covariance of Maxwell's equations and perform Lorentz transformation of (\ref{ezero}) from the original reference frame to the frame in which the momentum equals $p$. As in the case of Dirac equation, it is sufficient to use $SL(2,C)/E_2$ transformations because the little Wigner group $E_2$ leaves momentum invariant. We obtain
\be
\epsilon_{-}^{\mu}(p,r')=  \frac{1}{\sqrt{2}}\frac{\bar r_{\dot\alpha}'(\bar\sigma^{\mu})^{\dot\alpha\beta}\lambda_{\beta}}{[\lambda r']} ~,\qquad \epsilon_{+}^{\mu}(p,r')=\frac{1}{\sqrt{2}}\frac{r^{\prime\alpha}(\sigma^{\mu})_{\alpha\dot\beta}\bar \lambda^{\dot\beta}}{\langle r' \lambda\rangle}\label{epp}
\ee
where $r'$ denotes the Lorentz-transformed $r_0$.\bs
{\bf Exercise 5.3} Show that for arbitrary two spinors $r'$ and $r$,
\be
\epsilon_{\pm}^{\mu}(p,r')= \epsilon_{\pm}^{\mu}(p,r) +c_{\pm}p^\mu\ ,\label{egauge}
\ee
where
$$ c_+=c_-^*=\frac{\sqrt{2}\langle rr'\rangle}{\langle\lambda r\rangle\langle\lambda r'\rangle}$$\es\noindent
At the level of vector potential (\ref{ggf}), changing the polarization vector $\epsilon^{\mu}_s(p)\to \epsilon^{\mu}_s(p)+c_sp^\mu$ corresponds to
\be A^{\mu}_p(x)\to A^{\mu}_p(x)+\partial^\mu\Lambda(x)\ee
where
$$\Lambda(x)=i(c_sa_{\mathbf{p}}^se^{-ipx}-c_s^*a_{\mathbf{p}}^{\dag s}e^{ipx})\ .$$
  We see that the difference between vector potentials corresponding to polarization vectors $\epsilon_{\pm}^{\mu}(p,r')$ and  $\epsilon_{\pm}^{\mu}(p,r)$, c.f.\ Eq.(\ref{egauge}), amounts to a gauge transformation, therefore these fields are physically equivalent. We can use this fact and replace $r'_{\alpha}$ in Eq.(\ref{epp}) by arbitrary, fixed {\bf reference spinor} $r_{\alpha}$, the same for all momentum modes:
\be \epsilon_{-}^{\mu}(p,r)=  \frac{1}{\sqrt{2}}\frac{\bar r_{\dot\alpha}(\bar\sigma^{\mu})^{\dot\alpha\beta}\lambda_{\beta}}{[ \lambda r]} ~,\qquad \epsilon_{+}^{\mu}(p,r)=\frac{1}{\sqrt{2}}\frac{r^{\alpha}(\sigma^{\mu})_{\alpha\dot\beta}\bar \lambda^{\dot\beta}}{\langle r \lambda\rangle}\label{eppr}
\ee
The corresponding light-like vector $r^\mu=\frac{1}{2}r^{\alpha}(\sigma^{\mu})_{\alpha\dot\beta}\bar r^{\dot\beta}$ is called the {\bf reference vector}. Note that $r_\mu\epsilon^{\mu}_s(p,r)=0$.
This form of polarization vectors is the basic ingredient of the so-called {\bf spinor-helicity formalism} for evaluating Feynman diagrams which employs clever choices of reference spinors to simplify computations.\bs
{\bf Exercise 5.4} Show that
\be \sum_{s=\mp} \epsilon^{\mu}_s(p,r)\epsilon^{\nu *}_s(p,r)=-g^{\mu\nu}+\frac{p^\mu r^\nu+p^\nu r^\mu}{pr}\ee
\es
The quantum gauge field has the form
\be A^{\mu}(x)=\int\frac{d^3\pb}{(2\pi)^3}\frac{1}{\sqrt{2E_p}}\,\sum_{s=\mp}\big[\epsilon_s^{\mu}(p,r)e^{-ipx}
a_{\mathbf{p}}^s+\epsilon_{s}^{*\mu}(p,r)e^{ipx}
a_{\mathbf{p}}^{\dag s}\big]\ee
where $E_p=|\mathbf{p}|$ because $p^2=0$ for all modes. The creation operators create massless photons with two degrees of freedom. The respective helicities can be determined by rotating the wave functions by an angle $\theta$ about momentum direction which for a helicity $h$ eigenstate gives rise to the phase factor $e^{ih\theta}$ discussed in the context of Wigner's little group.
Since $\lambda\to e^{-i\theta/2}\lambda$ and $\bar\lambda\to e^{i\theta/2}\bar\lambda$ the photon wave functions (\ref{eppr}) $\epsilon_{\pm}^{\mu}(p,r)\to e^{\pm i\theta}\epsilon_{\pm}^{\mu}(p,r)$. Thus $h=\pm 1$.
The photon is a massless spin 1 particle in two possible helicity states, the left- and right-handed polarizations with  $h=-1$ and $h=1$, respectively.
\subsection{Lagrangian of QED}
In classical electrodynamics, charged particles are described by charge density and currents.
These are combined into one  relativistic current vector $j^\mu=(\rho,\mathbf{j})$ which is considered as the source of $F_{\mu\nu}$. In the presence of sources, Maxwell's equations read
\be \partial_{[\rho}F_{\mu\nu]}=0~~\mathbf{(M1:{\it dF=0})}\qquad \partial_{\mu}F^{\mu\nu}=j^\nu~~\mathbf{(M2':{\it \star \, d\star F=j})}\nonumber\label{meqs}\ .\ee
Note that $\mathbf{(M1)}$, which are Gauss' law of magnetism (absence of monopoles) and Faraday's law of induction, remain the same as without sources, therefore we can write  $F=dA$ as guaranteed by Stokes' theorem and use the vector potential $A^\mu$ as the primary field. Note that $\mathbf{(M2')}$ can only be consistent if the charge is conserved, i.e.\ it if the current satisfies the continuity equation $\partial_\mu j^\mu=0$.\footnote{Such a current is usually called the ``conserved current''.} In the Lagrangian density, the interaction between e-m fields and currents is described by a single, very simple term ${\cal L}_{\rm int}\sim A_\mu j^\mu$ and gauge invariance is guaranteed by charge conservation.

In QED, the problem boils down to constructing a quantum field theory of electrons and positrons interacting with photons, that is coupling Dirac theory to gauge fields. The Lagrangian density describing a free Dirac field is
\begin{equation}
    			\mathcal{L}_D=\bar{\psi}(i\gamma^\mu\partial_\mu-m)\psi\ .\label{dirl}
		\end{equation}\bs
{\bf Exercise 5.5} Show that the Lagrangian $\int d^4x \mathcal{L}_D$ is real, i.e.\ $\int d^4x \mathcal{L}_D^*=\int d^4x\mathcal{L}_D$\\
{\bf Exercise 5.6} Use variational principle to derive Dirac equation from $\mathcal{L}_D$.\\
{\bf Exercise 5.7} Construct the conserved Noether current $j^\mu$ associated to the invariance of $\mathcal{L}_D$ under $U(1)$ transformations $\psi(x)\to e^{i\alpha}\psi(x)$, where $\alpha$ is an arbitrary angle. This current is constructed up to an arbitrary normalization factor, but make sure that $j^{\mu\dag}=j^\mu$.\es
The QED Lagrangian describing Dirac fermions coupled to photons is constructed in the same way as in classical electrodynamics. The  $U(1)$ current is identified as the charge current and  it is coupled to the vector potential in the same way as in classical electrodynamics:
\be
\Lagr_{QED} =-\frac{1}{4} F_{\mu\nu}F^{\mu\nu} + \bar{\psi}(i \slashed{\partial} - m)\psi
~\underbrace{-~e A_\mu\, \bar{\psi}\gamma^{\mu}\psi\ .}_{\displaystyle\Lagr_{\rm int}~~~}
\label{lqed}\ee
Here $e$ is the electron charge.

The Maxwell Lagrangian describing free electromagnetic fields is invariant under gauge transformations
$A^{\mu}\to  A^{\mu} - \partial^{\mu}\alpha(x)$. This symmetry is preserved in QED once $\alpha(x)$ is identified with the (coordinate-dependent) angle parameter of the $U(1)$ symmetry responsible for electric charge conservation, and the vector field transformation is accompanied by a {\em local} $ U(1)$ transformation of charged fermions:
\be A^{\mu}\to  A^{\mu}{-}\,\partial^{\mu}\alpha(x)\ ,\quad
{\psi \to e^{ie\alpha(x)}\psi\atop\bar\psi \to  e^{-ie\alpha(x)}\bar\psi}\ .
\label{gtran}
\ee\bs
{\bf Exercise 5.8} Show that ${\cal L}_{QED}$ of Eq.(\ref{lqed}) is invariant under gauge transformations (\ref{gtran}).\\
{\bf Exercise 5.9} Rewrite  ${\cal L}_{QED}$ in two-component notation with
$\psi=\left(\chi_\alpha\atop\bar\eta^{\dot\alpha}\right)$ and show that it is invariant under Lorentz transformations.\es
\subsection{Non-Abelian Gauge Theory}
More interesting theories can be constructed for systems involving more than one species of Dirac fermions,
$\{\psi_i\}$, $i=1,2,\dots, n$:
\begin{equation}
    			\mathcal{L}_{D_n}=\sum_{i=1}^{n}\bar{\psi}^i(i\gamma^\mu\partial_\mu-m)\psi_i\ .\label{dirn}
		\end{equation}
In the standard model, $i=1,2,3$ labels three quark colors while a similar index $f=1,2$ labels electroweak doublets. A single quark doublet is described by $\Psi_{if}$ with $f$ labeling quark flavors, for instance up and down quarks. The Lagrangian (\ref{dirn}) is obviously invariant under
$\psi_i\to U_i^j\psi_j$
where $U_i^j$ are the elements of a $n\times n$ unitary matrix $U$ such that $U^\dag U=I$.  Hence $U(n)$ group, acting as $\psi\to U\psi,~\bar\psi\to\bar\psi U^\dag$, or any of its subgroups is a symmetry group of this Lagrangian. Here, we assume that the fermions transform in the fundamental representation of $U(n)$ although it is not difficult to generalize the following construction to other representations and/or groups. A $n\times n$ unitary matrix can be written as
$U=e^{i \alpha_a T^a}$, where $T^a$ are the hermitean generators and the real ``angles'' $\alpha_a, a=1,2,\dots n^2$ parameterize the transformation. One can also restrict to $SU(n)$ subgroup with
$n^2-1$ traceless generators.
For electro-weak $SU(2)$, $T^a=\frac{1}{2}\sigma^a$ while for $SU(3)$ QCD,
$T^a=\frac{1}{2}\lambda^a$, where $\lambda^a$ are the Gell-Mann matrices. The generators are always normalized as \be {\rm Tr}(T^aT^b)=\frac{1}{2}\delta^{ab}\label{tnorm}\ee and satisfy the commutation relations \be [T^a,T^b]=if^{abc}T^c,\label{tcom}\ee where $f_{abc}$ are called the structure constants. The currents associated to $U(n)$ invariance are
\be j^a= \bar{\psi}^i\gamma^{\mu}(T^a)^j_i\psi_j\ee

The QED Lagrangian is now generalized by introducing matrix-valued vector potentials $(A_\mu)_i^j$ and coupling them to the conserved currents as
\be
\Lagr_{QCD} =-\frac{1}{2g^2} {\rm Tr}(F_{\mu\nu}F^{\mu\nu}) + \bar{\psi}^i(i \slashed{\partial} - m)\psi_i
~\underbrace{-~\bar{\psi}^i\gamma^{\mu}(A_\mu)_i^j\psi_j}_{\displaystyle\Lagr_{\rm int}~~~}
\label{lqcd1}\ee
$(A_\mu)_i^j$ are hermitean matrices because $\Lagr_{\rm int}$ must be real. They can be expressed in terms of $n^2$ fields $A_\mu^a$ as
\be (A_\mu)_i^j=A_{\mu}^a(T^a)_i^j \ee
In order for ${\cal L}_{\rm int}$ to preserve (global) $U(n)$ invariance,  the potentials must transform as $A\to UAU^\dag$, i.e.\ in the $n^2$-dimensional adjoint representation of $U(n)$. Note that in Eq.(\ref{lqcd1}) we used a different normalization than in QED: we absorbed the coupling constant, now called $g$ instead of $e$, into the vector potential. Indeed, if we write $(F_{\mu\nu})^j_i=F_{\mu\nu}^a(T^a)_i^j$, then
\be\Lagr_{QCD} =-\frac{1}{4g^2} F_{\mu\nu}^aF^{a\mu\nu} + \bar{\psi}^i(i \slashed{\partial} - m)\psi_i
~\underbrace{-~\bar{\psi}^i\gamma^{\mu}A_\mu^a(T^a)_i^j\psi_j}_{\displaystyle\Lagr_{\rm int}~~~}
\label{lqcdd}\ee
which for the $U(1)$ field $A^0$ associated to $T^0=I/\sqrt{2n}$ has the same form as in QED with $n$ electrons provided that $A^0\to gA^0$ and $g\to e\sqrt{2n}$. In the next step, we want to generalize {\bf local} gauge invariance to the full $U(n)$ group, with the fermions transforming as
\be \psi \to U(x)\psi \qquad\bar\psi \to  \bar\psi U^\dag(x)~,\qquad U(x)=e^{i\alpha^a\!(x)\, T^a}\ .\label{psig}\ee
\bs {\bf Exercise 5.10} Show that ${\cal L}_{D_n}+{\cal L}_{\rm int}$ is invariant under the above transformation provided that the  $n\times n$ matrices $A_\mu$ transform as
\be A_\mu\to UA_\mu U^\dag+i\,U^\dag\partial_\mu U\ .\label{agt}\ee
Write the corresponding infinitesimal transformation (linear in $\alpha^a$) for the gauge fields $A_\mu^a$ (using the structure constants $f_{abc}$).
\es
This is not the end of the story. With the gauge field transforming as in Eq.(\ref{agt}), the Maxwell part of the Lagrangian constructed by using naive $F=dA$ field strength tensor is not gauge invariant. The problem is solved by introducing the ``non-abelian'' a.k.a.\ ``Yang-Mills'' field strength tensor ($n\times n$ matrix):
\be F_{\mu\nu}=\partial_\mu A_\nu-\partial_\nu A_\mu+ i[A_\mu,A_\nu]\ .\label{nfdef}
\ee\bs
{\bf Exercise 5.11} Show that under gauge transformations (\ref{agt})
\be F_{\mu\nu}\to UF_{\mu\nu}U^\dag\label{fgtran}\ee
and that the full Lagrangian $\int d^4x{\cal L}_{QCD}$ is invariant under gauge transformations
(\ref{psig}), (\ref{agt}).\\
{\bf Exercise 5.12} Write Eq.(\ref{nfdef}) as a definition of $F_{\mu\nu}^a$ in terms of $A_\mu^a$ (using the structure constants $f_{abc}$).\\
{\bf Exercise 5.13} Define the covariant derivative $D_\mu\psi=(\partial_\mu+iA_\mu)\psi$.  Show that under gauge transformations
$$D_\mu\psi\to UD_\mu\psi\ .$$\es\noindent
The QCD (or any non-abelian gauge theory with fermionic matter in the fundamental representations) Lagrangian density can be then written in a compact form as
\be\Lagr_{QCD} =-\frac{1}{4g^2} F_{\mu\nu}^aF^{a\mu\nu} + \bar{\psi}^i(i \slashed{D} - m)\psi_i\ .
\label{lqcd}\ee
\renewcommand{\refname}{\large Recommended Reading for Section 5}
  
\section{Primitive Amplitudes}\setcounter{equation}{0}
\subsection{Invariant Matrix Elements}
 We will be studying transition amplitudes describing the processes in which $n$ initial particles are scattered into $m$ final particles:
 \be
S_{fi}=\langle\mathbf{in}\,p_1,\dots, p_n|p_{n+1},\dots, p_{n+m}\,\mathbf{out}\rangle\ . \label{smel}\ee
All together, there are $m+n=N$ external particles participating in the process, so the corresponding amplitude will be called an $N$-particle amplitude. Particles can be ``moved'' from initial to final states by using ``crossing symmetries''. As far as the S-matrix are concerned, an incident particle with momentum $p$ and helicity $h$ is equivalent to an outgoing antiparticle with momentum $-p$ and helicity $-h$. Thus we can assume that all momenta $p_i$, $i=1,2,\dots, N$ are {\bf incoming}. With this convention, the momentum conservation law reads:
\be\sum_{i=1}^Np_i=0\ .\ee\bs
{\bf Exercise 6.1} Consider a scattering process of $N$ massless particles with $p_i^2=0$. Let $\lambda_l$ and $\bar\lambda_k$ be two arbitrary spinors. Show that the momentum conservation law can be expressed as
\be \sum_{i=1}^N\langle li\rangle[ik]=0\ .\ee\es
\noindent The {\bf S-matrix} elements (\ref{smel}) depend on the momenta, helicities and internal quantum numbers\footnote{We will use $a$ and other lower case latin letters to denote internal quantum numbers} of initial and final particles. The in and out states are built from single-particles states
\be |p,h,a\rangle = \sqrt{2E_{p}}\,a_{\mathbf{p}}^{\,\dag\, h,\, a}|0\rangle\quad,\qquad
\ . \label{ket}\ee
which are normalized as
\be\langle p',h',a' |p,h,a\rangle = 2E_{p}(2\pi)^3\,\delta_{h'h}\,\delta_{a'a}\,\delta^{(3)}(\mathbf{p'}-\mathbf{p}).\label{braket}\ee
All particles are on mass-shell,  $p^2=m^2$, hence $E_{p}=\sqrt{\mathbf{p}^2+m^2}$.

The S-matrix elements always contain momentum-conserving delta function factors, so it is convenient to factorize  them as
\be S_{fi}=i(2\pi)^4\delta^{(4)}\big(\sum_{i=1}^Np_i\big)\, {\cal M}(p_i,h_i,a_i),\label{mat}\ee
where $\cal M$ are the so-called ``invariant matrix elements'' or ``scattering amplitudes''.
\bs
{\bf Exercise 6.2} By using Eqs.(\ref{smel}, \ref{braket}, \ref{mat}), show that in mass units, in which $dim[p]=1$ and $dim[\lambda_p]=1/2$,
\be\dim[{\cal M}]=4-N\ ,\label{mdim}\ee
where $N$ is the  number of external (incoming plus outgoing) particles.
\es
\noindent A very important property of the amplitudes, which follows from (\ref{smel}) is the (anti)\linebreak symmetry under exchanging identical particles
\be{\cal M}(\dots i,\dots, j\dots)=\pm{\cal M}(\dots j,\dots, i\dots)\label{symm}\ee
with plus sign if $\{i,j\}$ are bosons and minus sign if they are fermions.

The amplitudes are Lorentz invariant. We will be mostly considering processes involving massless particles and expressing them in terms of spinor variables:
$${\cal M}(p_i,h_i,a_i)={\cal M}(\lambda_i,\bar\lambda_i,a_i)$$
In many cases, it will be convenient to convert some products $\langle ij\rangle[ji]$ into the Mandelstam-like invariants
$s_{ij}\equiv 2p_ip_j$.
\subsection{Feynman Rules}
If the interaction hamiltonian involves small coupling constants, as is fortunately in the case of standard model, the invariant matrix elements (\ref{mat}) can be computed perturbatively.
$i{\cal M}$ is usually computed by using Feynman diagrams constructed according to the Feynman rules.
The Feynman rules for QED and QCD are listed in the textbook, so there is no need of copying them here. We limit ourselves to listing external wavefunction factors associated to the particles of interest:
\begin{itemize}
\item For each incoming gauge boson: $\epsilon^{\mu}_s(p,r)$, with $s=\pm$ depending on helicity. An arbitrary reference $r$ can be chosen for each particle. There is also external gauge index $a$ labeling members of the adjoint multiplet.
\item For each incoming left-handed massless Weyl fermion: $\lambda_{\alpha}(\mathbf{p})$.
There is also external gauge index $i$ or $\bar\imath$ labeling  members of the fundamental or anti-fundamental representation, respectively, for quarks or electro-weak doublets.
\item For each incoming right-handed massless Weyl fermion: $\bar\lambda^{\dot\alpha}(\mathbf{p})$. Gauge index as above.
\item For each incoming Dirac fermion (electron or quark): $u^s(\mathbf{p})$ with $s=\pm$ depending on the spin state. Gauge index as above.
 \item For each incoming Dirac anti-fermion (positron or anti-quark): $\bar v^s(\mathbf{p})$ with $s=\pm$ depending on the spin state. Gauge index as above.
\end{itemize}
The most obvious, basic difference between non-abelian (Yang-Mills) gauge theory and QED is the presence of self-interactions among gauge bosons, visible  in three- and four-particle couplings. Unlike photons, non-abelian gauge bosons are ``charged'' and interact with each other. Even in the absence of matter, Yang-Mills gauge theory is an interacting theory with complex dynamics featuring asymptotic freedom and the existence of non-perturbative mass gap. QCD which describes quarks coupled to gluons is asymptotically free at short distances while at long distances it becomes strongly interacting, strong enough to confine quarks inside baryons and mesons.
\subsection{On-Shell Amplitudes for Three Massless Particles}
Feynman vertices represent interactions of three or more particles hence the most \linebreak ``primi{\nolinebreak}tive'' tree amplitudes
involve three external particles. Unlike Feynman vertices, the amplitudes are computed for physical, on mass-shell particles.
As we will see below, the three-particle kinematics are very restrictive. In order to relax these constraints we will be considering complex-valued momentum vectors.

Analytic continuation is a standard tool in theoretical physics. We will be considering complex-valued momenta for two reasons. First, in order to avoid kinematic restrictions for 3-particle processes and then, for complex contour integrations. In the first case, the problem is that in the simplest process of 3-particle scattering all kinematic invariants are trivial. In particular, in a process involving 3 massless particles all scalar products $2p_ip_j=\langle ij\rangle[ji]=0$, therefore $\langle ij\rangle=[ji]^*=0$. This reflects the momentum conservation law which is forcing these particles to move along a single line, hence their momenta are necessarily ``collinear''.

As explained before, momenta (real or complex, arbitrary length$^2=p^2$), can be represented by matrices
 \be p_{\alpha\dot\beta}\equiv p_\mu(\sigma^\mu)_{\alpha\dot\beta}\ ,\qquad\bar p^{\dot\alpha\beta}\equiv p_\mu(\bar\sigma^\mu)^{\dot\alpha\beta}\ .\ee
 More explicitly,
\be p=\left(\begin{array}{cc}
p_0+p_3 &p_1-ip_2\\ p_1+ip_2 &p_0-p_3\end{array}\right)\quad
\bar p=\left(\begin{array}{cc}
p^0+p^3 &p^1-ip^2\\ p^1+ip^2 &p^0-p^3\end{array}\right)\ .  \label{pmatrr}\ee
so that $\det p=\det\bar p=p^2$ is invariant under
Lorentz transformations.
{}For real momenta, $p$ and $\bar p$ are hermitean. Under the Lorentz group $SO(1,3)\sim SL(2,C)$, the momentum matrices transform as
\be   p\to M\,  p\,  M^{\dagger}\ ,\qquad \bar p\to M^{\dagger -1}\bar p\, M^{-1}\ . \label{pltran}   \ee
Note that the presence of $M^\dag$ is dictated by hermiticity. For light-like vectors with $\det p=0$, the matrices factorize as
$$ p_{\alpha\dot\beta}=\lambda_{\alpha}\bar{\lambda}_{\dot\beta}.$$
Here again, ${\lambda}_{\dot\beta}=(\lambda_{\beta})^*$ because of hermicity. Wigner's little group which leaves momentum invariant acts as $\lambda_{\alpha}\to e^{-i\theta/2}\lambda_{\alpha}$,
${\bar\lambda}_{\dot\beta}\to e^{i\theta/2}{\bar\lambda}_{\dot\beta}$.

Next, consider complex momentum vectors $p^\mu$. The most general transformation that leaves $\det p$ invariant is the similarity transformation
\be   p\to M\,  p\,  \widetilde{M}^{-1}\ ,\qquad \bar p\to \widetilde{M}\bar p\, M^{-1}\label{ptran},\ee
where $M$ and $\widetilde{M}$ are two arbitrary  $SL(2,C)$  matrices, hence the Lorentz group is extended to  $SL(2,C)\otimes \widetilde{SL}(2,C)$. For light-like complex momenta, it follows from Exercise 5.1 that
\be p_{\alpha\dot\beta}=\lambda_{\alpha}\tilde{\lambda}_{\dot\beta},\label{pfact}\ee
where $\lambda$ and $\tilde\lambda$ are two independent complex spinors which transform as
$$\lambda_\alpha\to(M\lambda)_\alpha,\quad \tilde\lambda^{\dot\alpha}\to(\widetilde M\tilde\lambda )^{\dot\alpha}.$$
Momenta remain unchanged under little group transformations
$$\lambda_\alpha\to t\lambda_\alpha,\quad \tilde\lambda^{\dot\alpha}\to t^{-1}\tilde\lambda^{\dot\alpha}$$
where $t$ is an arbitrary complex number replacing the previous $e^{-i\theta/2}$.
When the amplitudes are extended to complex momenta, $\tilde\lambda$ will replace $\bar\lambda$, also in square brackets which are no longer related by complex conjugation to angle brackets. Nevertheless,
$\langle ij\rangle[ji]=2p_ip_j$ remains true.

Wigner's little group provides a powerful constraint on the scattering amplitudes. Since the explicit momentum-spinor dependence of Feynman diagrams comes from external wave-function factors only, the amplitudes scale as
\be {\cal M}(t_i\lambda_i,(t_i)^{-1}\tilde\lambda_i,a_i)=\big(\prod_{i=1}^{N}t_i^{-2h_i}\big)\,{\cal M}(\lambda_i,\tilde\lambda_i,a_i)\label{wigs}\ee
We will be often using the above scaling together with ${\rm dim}{\cal M}=4-N$, that is with
\be c^d{\cal M}(c\lambda_i,c\tilde\lambda_i,a_i)=c^{2(4-N)}\,{\cal M}(\lambda_i,\tilde\lambda_i,a_i).\label{sca}\ee
where the factor $c^d$ comes from (eventual) dimensionful coupling constants. We will be also using Bose (or Fermi) symmetry properties of the amplitudes.

In the case of three massless particles, the momentum-dependence of the amplitudes can be determined by using little group scaling and dimensional analysis, without referring to Feynman diagrams.
We have on-shell conditions (OS) $p_ip_j=0$ and momentum conservation (MC) $p_1+p_2+p_3=0$, which are equivalent to
$$ \langle 12\rangle[12]=\langle 23\rangle[23]=\langle 31\rangle[31]=0 \qquad \bf (OS)$$
$$ \langle 12\rangle[23]=\langle 13\rangle[32]=\langle 23\rangle[31]=0 \qquad \bf (MC)$$
We can start from $[12]=0$ and $[23]=0$. Then automatically $[13]=0$ because the previous conditions imply
$1\parallel 3$ (in the sense of spinors being proportional to each other). Now (OS) and (MS) are satisfied for arbitrary angle products. We call this solution ``holomorphic''. In a similar way, we obtain ``antiholomorphic'' solution with $\langle 12\rangle=\langle 23\rangle=\langle 13\rangle=0$ and arbitrary square products. On the other hand, if we start with $[12]=0$ and $\langle 23\rangle=0$, we must also satisfy
$\langle 31\rangle[31]=0$ which leads to a special case of holomorphic or antiholomorphic solution. So, up to numerical and group factors, we have two candidates for three-particle amplitudes:
\be {\cal M}(h_1,h_2,h_3)=\langle 12\rangle^{d_3}\langle 23\rangle^{d_1}\langle 31\rangle^{d_2}\ ,\qquad \widetilde{\cal M}(h_1,h_2,h_3)=[12]^{\tilde d_3}[23]^{\tilde d_1}[31]^{\tilde d_2},\ee
the holomorphic and antiholomorphic ones, respectively. Next, we use little group scaling (\ref{wigs}) which yields
$$d_2+d_3=-2h_1,~d_1+d_3=-2h_2, ~d_1+d_2=-2h_3$$ $$\boldsymbol{\Longrightarrow} ~~d_1=h_1-h_2-h_3,~
d_2=h_2-h_1-h_3,~d_3=h_3-h_1-h_2$$
Obviously, $\tilde d_i=-d_i$.

As an example, we discuss three gluons, starting from the holomorphic case, with the following $\pm 1$ helicity patterns:
$${\cal M}(+,+,+)=\frac{1}{\langle 12\rangle\langle 23\rangle\langle 31\rangle}\ ,\quad
{\cal M}(-,-,-)=\langle 12\rangle\langle 23\rangle\langle 31\rangle$$
$${\cal M}(+,+,-)=\frac{\langle 23\rangle\langle 31\rangle}{\langle 12\rangle^3}\ ,\qquad
{\cal M}(-,-,+)=\frac{\langle 12\rangle^{3}}{\langle 23\rangle\langle 31\rangle}$$
Three-particle amplitude has dimension 1, c.f.\ Eq.(\ref{mdim}). The gauge coupling constant is dimensionless, therefore only one holomorphic amplitude, ${\cal M}(-,-,+)$,
has the right dimension. Among antiholomorphic amplitudes, only $\widetilde{\cal M}(+,+,-)$ is allowed on similar grounds. Thus we expect
\be {\cal M}(+,+,+)=
{\cal M}(-,-,-)=0\label{mhv0}\ee \be
{\cal M}(-,-,+)=\frac{\langle 12\rangle^{3}}{\langle 23\rangle\langle 31\rangle}~,\qquad
{\cal M}(+,+,-)= \frac{[12]^{3}}{[ 23][ 31]}\label{mhv3}\ee
But this is not the end of the story. The amplitudes (\ref{mhv3}), as written above, are antisymmetric under exchanging identical particles $1\leftrightarrow 2$. Gluons are bosons therefore the amplitude should be symmetric. This is not a problem. Until this point, we ignored group factors depending on the group indices of three gluons. This factor should be a completely antisymmetric group invariant, but the only one available is the group structure constant. Hence
\be
{\cal M}(-, a;-, b;+,c)=x\, f_{abc}\,\frac{\langle 12\rangle^{3}}{\langle 23\rangle\langle 31\rangle}~,\quad
{\cal M}(+,a;+,b;-,c)= \tilde x \,f_{abc}\,\frac{[12]^{3}}{[ 23][ 31]}\label{mhv33}\ee
where $x$ and $\tilde x$ are overall constants that can be only determined by an explicit computation.

Did we get is right? The only way to verify our conclusions is by performing a Feynman diagram computation. In this case, there is only one diagram, with external wave-functions directly attached to the three-gluon vertex. Recall that
 \be \epsilon_{-}^{\mu}(p,r)=  \frac{1}{\sqrt{2}}\frac{\tilde r_{\dot\alpha}(\bar\sigma^{\mu})^{\dot\alpha\beta}\lambda_{\beta}}{[ \lambda r]} ~,\qquad \epsilon_{+}^{\mu}(p,r)=\frac{1}{\sqrt{2}}\frac{r^{\alpha}(\sigma^{\mu})_{\alpha\dot\beta}\tilde \lambda^{\dot\beta}}{\langle r \lambda\rangle}\ .
\ee
 We can use three different reference vectors, $r_1,r_2,r_3,$ for each polarization vector. Due to the presence of $g_{\mu\nu}$ in the vertex, all contributions contain factors $\epsilon_i\epsilon_j$, of two polarization vectors contracted with each other.  If all polarizations are identical, (+++) or $(---)$, we can choose all reference vectors equal, $r_1=r_2=r_3$. Then all $\epsilon_i\epsilon_j=0$ and we prove Eq.(\ref{mhv0}). Next, we consider the case of $(--+)$. The choice of $r_1=r_2=r$ and $r_3=p_1$ leaves only one nonvanishing contraction (for the moment, we ignore numerical factors),
 \be\epsilon_2\epsilon_3\approx \frac{\langle 12\rangle [3r]}{[2r]\langle 13\rangle}\label{eeps}\ee
 The amplitude has the form
  $$(\epsilon_2\cdot \epsilon_3)\times\epsilon_1\cdot (p_2-p_3)= 2(\epsilon_2\cdot \epsilon_3)(\epsilon_1\cdot p_2)\approx\frac{\langle 12\rangle^2 [3r]}{\langle 31\rangle[1r]} \ ,$$
 where we used momentum conservation $p_3=-p_1-p_2$ and transversality $\epsilon_1p_1=0$.
 Next, we multiply numerator and denominator by $\langle 23\rangle$ and use momentum conservation in $\langle 23\rangle [3r]=
\langle 12\rangle [1r] $, so that
$$\frac{\langle 12\rangle^2 [3r]}{\langle 31\rangle[1r]}=\frac{\langle 12\rangle^3}{\langle 23\rangle\langle 31\rangle}\ .$$
After inserting the group factor, we reproduce the $(--+)$ part of Eq.(\ref{mhv33}).\bs
{\bf Exercise 6.3} Repeat the computation of $(+++),~(---),~(--+),~(++-)$ amplitudes by using generic
$r_1,r_2,r_3$ reference vectors. Show that results do not depend on the choice of these vectors and determine the constants $x$ and $\tilde x$ in Eq.(\ref{mhv33}).\\
{\bf Exercise 6.4} An alternative way of relaxing kinematic constraints for 3-particles processes is by changing the spacetime metric signature to $(+,-,+,-)$, which can be accomplished by $p_2\to ip_2$. Now the Lorentz group is $SO(2,2)$. Show that the momentum matrix (\ref{pmatrr}) can be factorized as in (\ref{pfact}) with real $\lambda$ and $\tilde\lambda$. How does the Lorentz group act on these spinors?\es
\renewcommand{\refname}{\large Recommended Reading for Section 6}
  
\section{Structure of Tree Amplitudes}\setcounter{equation}{0}
\subsection{Group Factors}
Each Feynman diagram contributing to the amplitude comes with its own group factor depending on group indices of external particles. It is very important to develop a systematic method of managing these factors. The basic observation is that any $N$-gluon amplitude can be written as
$$ {\cal M}(\{p_1,h_1,a_1\},\{p_2,h_2,a_2\},\{p_3,h_3,a_3\},\dots,\{ p_n, h_N ,a_N\})=~~~~~~~~~~~~~~~~~$$
\be \sum_{\sigma\in S_{N{-}1}}2^{N/2}\,{\rm tr}\big(T^{a_1}T^{a_{2\sigma}}T^{a_{3\sigma}}\cdots T^{a_{N\sigma}}\!\big)\, A\big (1^{h_1},2_{\sigma}^{h_{2\sigma}},3_\sigma^{h_{3\sigma}},\dots,N_\sigma^{h_{N\sigma}}\big)\label{partial1}\ee
where the sum is over $(N-1)!$  permutations $\sigma$ of the set $\{2,3,4,\dots,N\}$ and $i_\sigma\equiv \sigma(i)$. Superscripts $h_{i\sigma}$ keep track of particle helicities. The coefficients\\ $A\big (1^{h_1},2_{\sigma}^{h_{2\sigma}},3_\sigma^{h_{3\sigma}},\dots,N_\sigma^{h_{N\sigma}}\big)$ are called partial amplitudes.\footnote{The overall factor $2^{N/2}$ was introduced in order to keep the standard group-theoretical normalizations (\ref{tnorm}) without changing the conventional normalization of partial amplitudes} These partial amplitudes (sometimes caller ``color-stripped'') do not depend on group indices. They depend of kinematic invariants (i.e.\ on vector and spinor products) only and are {\em universal for all gauge groups}. As an example,
$$ {\cal M}(\{p_1,-,a_1\},\{p_2,-,a_2\},\{p_3,+,a_3\})=~~~~~~~~~~~~~~~~~~~~~~~~~~~~~~~~~~~$$
\be 2^{3/2}{\rm tr}\big(T^{a_1}T^{a_{2}}T^{a_{3}}\big) A\big (1^-2^-3^+)
+2^{3/2}{\rm tr}\big(T^{a_1}T^{a_{3}}T^{a_{2}}\big) A\big (1^-3^+2^-\big)\ .\label{thrp}\ee

The proof of Eq.(\ref{partial1}) is very simple. In any tree level Feynman
diagram, replace the group factor at some vertex using $f_{abc} = -2i {\rm tr}(T^aT^bT^c-T^cT^bT^a)$.
Now each leg attached to this vertex has a $T$ matrix associated with it. At the other end of each of
these legs there is either another vertex or this is an external leg. If there is another vertex, connected say to $c$, we use the
$T^c$ associated with this internal leg to write the color structure of this vertex $f_{cde}T^c$  as $-i [T^d, T^e]$.
Continue this processes until all vertices have been treated in this manner. Then the group factor of the Feynman
diagram has been placed in the form of a trace of $N$ generators labeled by group indices of external particles. By using the cyclic property of the trace, we can always keep $T^{a_1}$ as the first factor, so this Feynman diagram has been placed in the form of Eq.(\ref{partial1}). Repeating this procedure for all Feynman
diagrams for a given process completes the proof.
\bs {\bf Exercise 7.1} Use results of Exercise 6.3 in order to determine $A(1^-2^-3^+)$ and $A(1^-3^+2^-)$\es

Let's discuss this color decomposition for four-gluon scattering. There are four diagrams, shown on Fig.1.
\begin{figure}[h!]
\begin{center}
\includegraphics[scale=2.5]{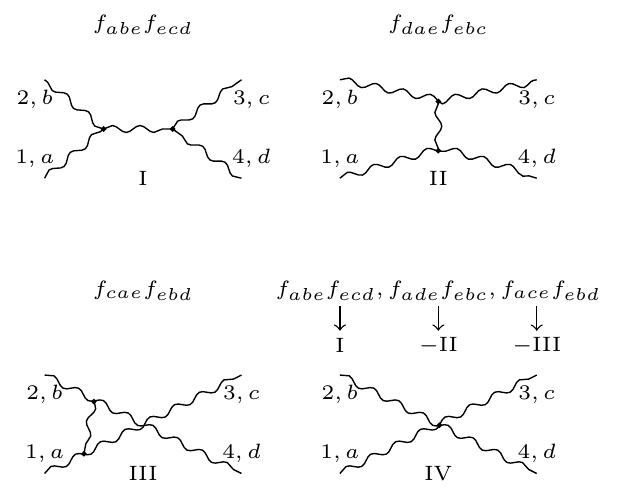}
\end{center}
\caption{Feynman diagrams and the corresponding group factors for four-gluon scattering}
\label{rf1}
\end{figure}
The group factor of diagram IV with the four-gluon vertex is a linear combinations of diagrams I, II and III. Thus as far as color structure is concerned, there is no need to consider four-gluon vertices because their effect is always equivalent to a sum of 3 three-gluon vertices. Let's follow the procedure described in the proof:
$$ {\rm I}:~ f_{abe}f_{ecd}\approx {\rm tr}\big([T^a,T^b][T^c,T^d]\big)\ ,\qquad
{\rm II}:~ f_{dae}f_{ebc}\approx {\rm tr}\big([T^d,T^a][T^b,T^c]\big)$$
$$ {\rm III}:~ f_{cae}f_{ebd}\approx {\rm tr}\big([T^c,T^a][T^b,T^d]\big)$$
Suppose that we want to determine the partial amplitude $A(1,2,3,4)$, that is the coefficient of
${\rm tr}\big(T^aT^bT^cT^d\big)$. We see that we have to combine the  kinematic parts of Feynman diagrams as
$\rm I+II$ (plus of course the relevant part of IV). Note that III does not contribute to this partial amplitude.\bs
{\bf Exercise 7.2} Compute $A\big (1^-2^+3^-4^+)$. Show that this amplitude can be written as a ``holomorphic'' rational function of angle products. Hint: A smart choice of reference vectors will simplify the computation enormously. Here I suggest $r_1=r_3=p_2$, $r_2=r_4=p_3$. This will set many $\epsilon_i\cdot\epsilon_j$ to zero.\\
{\bf Exercise 7.3} Show that ${\cal M}(\{p_1,+,a_1\},\{p_2,+,a_2\},\{p_3,+,a_3\}),\{p_4,+,a_4\})=$\\ ${\cal M}(\{p_1,+,a_1\},\{p_2,+,a_2\},\{p_3,+,a_3\}),\{p_4,-,a_4\})=0$, and the same thing\\ with $+\leftrightarrow -$.
One way to state this result is that at the tree level all plus amplitudes and three plus one minus amplitudes are vanishing. Hint: A smart choice of reference vectors will simplify the computation enormously.\es
\subsection{Properties of Partial Amplitudes}

~~~~{\bf A) Partial amplitudes are gauge invariant}\\[1mm]
Proof: The full amplitude on the l.h.s.\ of (\ref{partial1}) is gauge invariant, so the question is whether a variation of one partial amplitude can be canceled by another one. This could be possible if the group factors were linearly dependent, with the same linear combinations vanishing for all groups. Let's start from 3-gluon amplitude, see Eq.(\ref{thrp}). Begin with SU(2). In this case,
$${\rm tr}(T^{a}T^{b}T^{c})=\frac{1}{2}{\rm tr}\big(T^a\{T^{b},T^{c}\})+\frac{1}{2}{\rm tr}\big(T^a[T^{b},T^{c}])=\frac{i}{4}f_{abc}\qquad {\bf SU(2)}\ ,$$
so $\big(T^{a_1}T^{a_{2}}T^{a_{3}}\big)=-{\rm tr}\big(T^{a_1}T^{a_{3}}T^{a_{2}}\big)$. But this is peculiar for SU(2). For SU(n) with $n>2$
$${\rm tr}(T^{a}T^{b}T^{c})=\frac{1}{4}(d_{abc}+if_{abc}) \qquad {\bf SU(n)}\ ,$$
where $d_{abc}$ are the totally symmetric SU(n) invariants. Let
$\delta_1=\delta A \big (1^-2^-3^+)$ be the gauge variation of $A \big (1^-2^-3^+)$ and similarly,
$\delta_2=\delta A\big (1^-3^+2^-\big)$. Then
$$0=\delta {\cal M} =\frac{1}{4}d_{abc}(\delta_1+\delta_2)+\frac{i}{4}f_{abc}(\delta_1-\delta_2),$$
hence $\delta_1=\delta_2=0$ because $d_{abc}$ and $f_{abc}$ are independent. With some more sophistication in group theory, one can prove in a similar way that the trace factors in Eq.(\ref{partial1}) are always linearly independent, in the sense explained above.\vskip 1mm

{\bf B) U(1) decoupling (Kleiss-Kuijf relation)}:
$$ A(1,2,3,\dots, N{-}1,N)+A(1,2,3,\dots, N,N{-}1)+A(1,2,3,N,\dots, N{-}1)+\dots$$\be \dots +A(1,2,N,3,\dots, N{-}1)+A(1,N,2,3,\dots, N{-}1)=0\ .\label{kkl}\ee\vskip 1mm
Proof: If one of gauge bosons in the tree amplitude describing the scattering of $N$ $U(n)$ gauge bosons is associated to the $T^0=I_n/\sqrt{2n}$ $U(1)$ subgroup generator of $U(n)$, the amplitude vanishes because $f_{0ab}=0$, or in other words, this abelian gauge boson does not couple to gluons. Let's assume that it is the $N^{\rm th}$ gauge boson. Then according to Eq.(\ref{partial1}),
 $$ 0=\frac{1}{\sqrt{2n}}\sum_{\sigma\in S_{N{-}2}}{\rm tr}\Big(T^{a_1}T^{a_{2\sigma}}T^{a_{3\sigma}}\cdots T^{a_{(N-1)_\sigma}}\!\Big)\, A\big (1,2_{\sigma},3_\sigma,\dots,(N{-1})_\sigma,N\big)~~~~~~~~~~~$$
$$ +\frac{1}{\sqrt{2n}}\sum_{\rho\in S_{N{-}2}}{\rm tr}\Big(T^{a_1}T^{a_{2\rho}}T^{a_{3\rho}}\cdots T^{a_{(N-1)_\rho}}\!\Big)\, A\big (1,2_{\rho},3_\rho,\dots,N,(N{-}1)_\rho\big)+\dots$$
\be\dots +~\frac{1}{\sqrt{2n}}\sum_{\pi\in S_{N{-}2}}{\rm tr}\Big(T^{a_1}T^{a_{2\pi}}T^{a_{3\pi}}\cdots T^{a_{(N-1)_\pi}}\!\Big)\, A\big (1,N,2_{\pi},3_\pi,\dots,(N{-1})_\pi\big).\ee
After choosing all permutations $\sigma=\rho=\dots=\pi={\rm id}$, and using linear independence of group traces, we obtain Eq.(\ref{kkl}). Note that while the derivation used $U(n)$, the result is valid for any gauge group!\vskip 1mm

{\bf C) (Reflection) Parity:}
\be A(1,2,3,4,\dots N)=(-1)^N A(1,N, N{-}1,\dots,2)\ .\label{parity1}\ee
\vskip 1 mm
Proof: This follows from the properties of Feynman diagrams. For $N=3$ it is obvious due to antisymmetry of $f_{abc}$. For higher $N$, Feynman diagrams are symmetric (even $N$) or antisymmetric (odd $N$) with respect to  mirror reflections which reverse the order of external gauge indices from clockwise to anticlockwise.
\vskip 2mm\noindent
\subsection{From Amplitudes to Cross Sections: Squares, Color Sums and Kinematics}
One of important applications of our formalism is to the $SU(3)$ QCD gluon (and quark) scattering processes at hadron colliders. These are the so-called hard parton sub-processes, responsible for the production of hadronic jets.
Here the object of interest is $|{\cal M}|^2$ {\em averaged} over initial spin and colors and {\em summed} over final spin and colors. We call it $\overline{|{\cal M}|^2}$. For the purpose of this section, we have to temporarily return to real momenta ($\tilde\lambda\to\bar\lambda)$, reverse the momentum sign for outgoing particles etc.

First, let us count the number of $i$'s in Feynman diagrams. It is easy to see, counting $i$'s from propagators and four-gluon vertices, that vertices and propagators (together with overall $i$ from $i{\cal M}$) yield $i^{N}$ multiplied by a real function of momenta, times external wave functions. We also know that $\epsilon^*_-=\epsilon_+$ and $\lambda^*=\bar\lambda$. Hence
$$ {\cal M}^*(\{p_1,h_1,a_1\},\{p_2,h_2,a_2\},\{p_3,h_3,a_3\},\dots,\{ p_n, h_N ,a_N\})=~~~~~~~~~~~~~~~~~~~~~$$
$$(-1)^N {\cal M}(\{p_1,-h_1,a_1\},\{p_2,-h_2,a_2\},\{p_3,-h_3,a_3\},\dots,\{ p_n, -h_N ,a_N\})\ .$$
Now let's compare it with complex conjugation applied directly to Eq.(\ref{partial1}). By using the hermiticity property of $T$ matrices,
\be {\rm tr}\big(T^{a_1}T^{a_{2}}T^{a_{3}}\cdots T^{a_{N}}\!\big)^*={\rm tr}\big(T^{a_N}T^{a_{N{-}1}}\cdots T^{a_{2}}T^{a_{1}}\!\big)\ ,\label{ttran}\ee
and reflection parity (\ref{parity1}), we find
\be A(1^{h_1},2^{h_2},\dots, (N{-}1)^{h_{N{-}1}},N^{h_N})^*=A(1^{-h_1},2^{-h_2},\dots, (N{-}1)^{-h_{N{-}1}},N^{-h_N})\ ,\ee
which is very useful when considering all helicity configurations.

After squaring Eq.(\ref{partial1}), we often need to perform sums over group indices, like
\be \sum_{a_1,a_2,\dots}{\rm tr}\big(T^{a_1}T^{a_{2}}T^{a_{3}}\cdots T^{a_{N}}\!\big){\rm tr}\big(T^{a_{N_\pi}}T^{a_{(N{-}1)_\pi}}\cdots T^{a_{2\pi}}T^{a_{1}}\!\big)\ ,\label{tsum}\ee
where $\pi$ are some permutations. Note that each index appears twice. At this point, it is convenient to use the completeness property of the basis of hermitean matrices:
\be \sum_{a=1}^{n^2}(T^{a})^i_j(T^{a})^k_l=\frac{1}{2}\delta^i_l\delta^k_j\qquad \qquad \bf U(n).\ee
If we want to sum over SU(n) indices only, like in QCD, then we subtract the U(1) generator:
\be \sum_{a=1}^{n^2-1}(T^{a})^i_j(T^{a})^k_l=\frac{1}{2}\delta^i_l\delta^k_j-\frac{1}{2n}\delta^i_j\delta^k_l\ \qquad \quad \bf SU(n)\ .\ee\bs
{\bf Exercise 7.4}. Show that for SU(n) or U(n) at large $n$,
 $$\sum_{a_1,a_2,\dots, a_N}{\rm tr}\big(T^{a_1}T^{a_{2}}T^{a_{3}}\cdots T^{a_{N}}\!\big){\rm tr}\big(T^{a_{1}}T^{a_{2\sigma}}T^{a_{3\sigma}}\cdots T^{a_{N\sigma}}\!\big)^*=\left(\frac{n}{2} \right)^N\delta({\sigma-{\rm id}})+{\cal O}(n^{-2}).$$
 so there is no interference between partial amplitudes in the large $n$ limit.\es

In our conventions, all momenta, helicities and charges are defined for incident particles. If we want to compute a cross section with a number of {\em outgoing} particles, we can revert the respective momenta from $p$ to $-p$ {\em after} computing $|{\cal M}|^2$. We should keep in mind that in this process, the respective helicities are also reverted and particles are replaced by anti-particles, as implied by CPT symmetry.
After using $|\langle ij\rangle|^2=2 p_ip_j$ and the momentum conservation law, since $|{\cal M}|^2$ is invariant under Wigner's little group, we can always eliminate all momentum spinors and express $|{\cal M}|^2$ in terms of Lorentz-invariant scalar products of momentum vectors. Hence there is no need for reverting the momentum sign for momentum spinors. {}For example in the process $1+2\to 3+4$, $|{\cal M}|^2=|{\cal M}|^2(s,t,u)$ with $s=2p_1p_2$, $t=-2p_1p_3$, and $u=-2p_2p_3$, where $p_1$ and $p_2$ are incident momenta while  $p_3$ and $p_4$ are the outgoing ones.

What is the number of variables necessary to determine all momenta in a given reference frame?
 All momenta are on shell, so we start with $3N$ components. The conservation law brings 4 constraints, and further 6 components can be fixed by using Lorentz transformations. We end up with $3N-10$ independent kinematic variables. For $N=4$, the standard choice are the Mandelstam's variables $s=2p_1p_2$ and $u=-2p_2p_3$, although one also customarily uses $t=-2p_1p_3$, remembering that $s+t+u=0$. For $N=5$, one can choose $s_1=2p_1p_2$, $s_2=2p_2p_3~,\dots,s_5=2p_5p_1$, etc.

 The final step is averaging over initial helicities and group indices. For a process with $q$ initial gauge bosons,
 $$\overline{|{\cal M}|^2}=\frac{1}{\big(2{\rm dim}{\cal G}\big)^{q}}\,\sum_{h_i,\, a_i}|{\cal M}|^2\ ,$$
 where $\rm dim\cal G$ is the dimension of Lie algebra, e.g. ${\rm dim}{\cal G}=n^2-1$ for $SU(n)$, and the factor of 2 comes from averaging over $h=\pm 1$ for each incident gluon.\bs
 {\bf Exercise 7.5}. Compute $\overline{|{\cal M}|^2}$ for the elastic scattering of two $SU(n)$ gluons,  $1+2\to 3+4$, and express it in terms of $n$ and Mandelstam's variables $s,t,u$. \es
 \renewcommand{\refname}{\large Recommended Reading for Section 7}

 \section{Soft and Collinear Limits}\setcounter{equation}{0}
\subsection{Soft Singularities}
 In Feynman diagrams, internal lines represent the propagators of virtual particles:
\be D^{\mu\nu}(p)=-i g^{\mu\nu}\frac{1}{p^2}\ee
for gauge bosons (in Feynman gauge) and
\be S_{\alpha\dot\beta}(p)=i p_{\alpha\dot\beta}\,\frac{1}{p^2}\ee
for Weyl fermions.
These propagators have poles
at $p^2=0$, reflecting the breakdown of perturbation theory when virtual particles approach the on-shell limit of
$p^2\to 0$, i.e. when single-particle (resonance) production channels open up as  dominant processes.\footnote{If the particle is unstable, then the on-shell singularity is ``smeared'' by quantum corrections, producing an observable ``resonance peak.''}

In the soft emission process, a virtual particle goes on shell by emitting a soft particle with momentum $k\approx 0$. The singularity appears because of the pole of the intermediate propagator. This soft singularity reflects the fact that every experiment has a definite energy/momentum resolution therefore it is not possible to distinguish a given scattering process from a process in which one or even a larger number of additional, very soft particles is produced. For example, $(N{+}1)$-gluon (squared) amplitude with one soft gluon should be interpreted as appropriately corrected (squared) $N$-gluon amplitude. The details are covered by Bloch-Nordsieck theorem; here we do not discuss the interpretation, but we focus on
the mathematical description of soft singularities, restricted to the case of $N$-gluon Yang-Mills amplitudes.

The singular contributions to a soft emission process arise from Feynman diagrams in which a virtual gluon goes on-shell by emitting a soft gluon. The splitting is due to the three-gluon vertex with the structure constant $f_{a_ka_na_I}$ linking the soft gluon $a_k$ and the final gluon $a_n$ together with the propagator connecting to an off-shell amplitude in which $T^{a_I}$ appears inside trace factors. Since
$f_{a_ka_na_I}T^{a_I}=-i[T^{a_k},T^{a_n}]$,
the group index $a_k$ of the soft  gluon must be adjacent to the index of the gluon $a_n$ in the corresponding contribution to the full amplitude in order to produce $(k+p_n)^2=2kp_n\to 0$ singularity. In other words, the $k\to 0$  singularity of each partial amplitude is determined by the gluons adjacent to $a_k$. Thus the singular behaviour of contribution associated to
$$2^{N/2}{\rm tr}(T^{a_1}T^{a_2}\cdots T^{a_m}T^{a_k}T^{a_n}\cdots T^{a_N} )A(1,2,\dots m,k,n,\dots, N),$$
is determined by the diagrams with $k$ attached to $m$ or $n$  via the three-gluon vertex, see Fig.2.
\begin{figure}[h!]
\begin{center}
\includegraphics[scale=1.3]{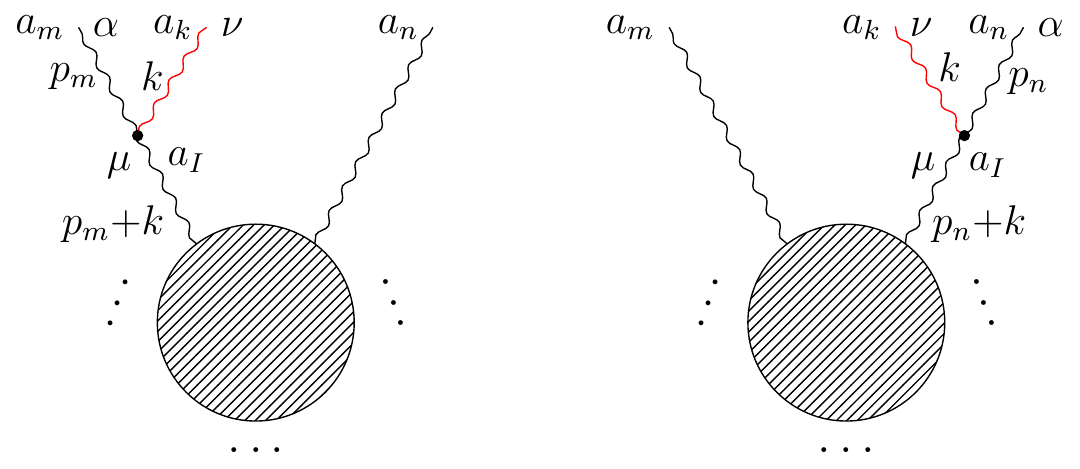}
\end{center}
\caption{Feynman diagrams contributing to $k\to 0$ soft singularity}
\label{rf2}
\end{figure}
Let us start from the $n$th gluon. The propagator and three-gluon vertex contribute
\be \frac{-i}{2kp_n}(-i)(-g)\Big[(\epsilon_k\epsilon_n)(k-p_n)^\mu-2(\epsilon_nk)
\epsilon_k^{\mu}+2(\epsilon_kp_n)\epsilon_n^{\mu}
\Big]\ee
where $\mu$ is the Lorentz index of the other end of the propagator, connected to the rest of the Feynman diagram. When $k\to 0$, $(k-p_n)^\mu\to -p^\mu_n$, therefore the first term in the square bracket vanishes by gauge invariance. The second term is already of ${\cal O}(k)$ order. Hence only the last term  contributes from the bracket at the leading ${\cal O}(k^0)$ order. Since the polarization vector $\epsilon_n^{\mu}$ is contacted with the rest of the diagram, the net effect is that, up to an overall factor, the $(k,n)$ pair is replaced by a single gluon carrying the momentum and helicity of the $n$th gluon. A similar contribution from the $m$th gluon comes with the opposite sign due to antisymmetry of $f_{abc}$. As a result, at the leading order
\be A(1,2,\dots m,k,n,\dots, N)\to \frac{g}{\sqrt 2}\left(\frac{\epsilon_kp_n}{kp_n}-\frac{\epsilon_kp_m}{kp_m}\right)
A(1,2,\dots m,n,\dots, N)\label{soft1}\ee
Note that the above expression is gauge invariant because the soft ``eikonal'' factor vanishes for $\epsilon_k=k$. This factor can be expressed in terms of spinor products in the following way:
\begin{eqnarray}A(1,2,\dots m,k^-,n,\dots, N)&\to& g\frac{[mn]}{[mk][kn]}
A(1,2,\dots m,n,\dots, N)\\
A(1,2,\dots m,k^+,n,\dots, N)&\to& g\frac{\langle mn\rangle}{\langle mk\rangle\langle kn\rangle}
A(1,2,\dots m,n,\dots, N)\end{eqnarray}\bs
{\bf Exercise 8.1} Derive the above result from Eq.(\ref{soft1}) and check all factors of $\sqrt 2$ and signs.\es
\noindent Note that if the soft momentum spinors scale as $\lambda_k\sim \epsilon$, then the singularity corresponds to a double pole $1/\epsilon^2$. Recently, a compact expression has been worked out for the sub-leading soft singularities corresponding to $1/\epsilon$ poles \cite{Casali:2014xpa}.
\subsection{Collinear Limit}. In general, for two massless particles $i$ and $\!j$, the collinear limit is defined as a special kinematic configuration in which the particles propagate with parallel four-momentum vectors, with the total momentum $P$ distributed as $p_m=xP$ and $p_n=(1-x)P$, so that $ (p_m+p_n)^2=P^2=0$.
This configuration gives rise to propagator poles in Feynman diagrams in which a virtual gluon splits into the collinear pair via three-gluon interactions.
 This singularity reflects a finite angular resolution of particle detectors which are unable to distinguish between one or more particles moving in the same direction. In this case, the Kinoshita-Lee-Nauenberg theorem gives a prescription for handling such singularities.

In order to give a precise definition of the leading and subleading parts of the amplitude, we need to specify how the collinear limit is reached from a generic kinematic configuration. Let us specify to generic light-like momenta $p_m,p_n$ and introduce two light-like vectors $P$ and $r$ such that the momentum spinors decompose as
$$\lambda_{m}=\lambda_P\cos\theta-\epsilon\lambda_r\sin\theta
\quad\quad\quad\tilde\lambda_{m}=\tilde\lambda_P\cos\theta
-\epsilon\tilde\lambda_r\sin\theta $$ $$
\lambda_{n}=\lambda_P\sin\theta+\epsilon\lambda_r\cos\theta
\quad\quad\quad~~~\,\tilde\lambda_{n}=\tilde\lambda_P\sin\theta+
\epsilon\tilde\lambda_r\cos\theta$$
hence
$$p_{m} ={\bf c}^2P-\epsilon\,{\bf s c} (\lambda_P\tilde\lambda_r+
\lambda_r\tilde\lambda_P)+ \epsilon^2{\bf s}^2 r $$ $$
p_{n}={\bf s}^2P+\epsilon\,{\bf sc} (\lambda_P\tilde\lambda_r+
\lambda_r\tilde\lambda_P)+ \epsilon^2{\bf c}^2 r
$$
where
$${\bf{c}}\equiv\cos\theta=\sqrt{x}~,\qquad\qquad {\bf{s}}\equiv\sin\theta=\sqrt{1-x}\ .$$
We also have
$$\langle mn\rangle=\epsilon\,\langle Pr\rangle~, \qquad
[mn]=\epsilon\,[Pr]~.$$
The total momentum
$$p_{m}+p_n=P+\epsilon^2r~,\qquad (p_{n}+p_n)^2= 2 Pr\,\epsilon^2\ .$$
The collinear configuration will be reached in the $\epsilon\to 0$ limit and
 the tree amplitudes  can be expanded in powers of $\epsilon$.
 \begin{figure}[h!]
\begin{center}
\includegraphics[scale=1.3]{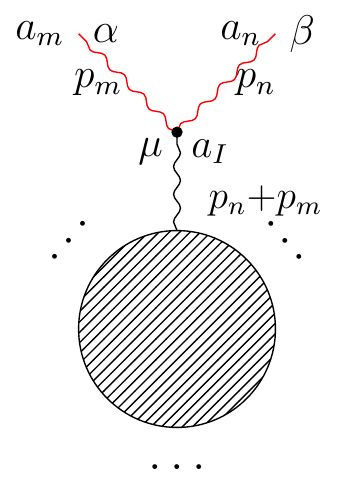}
\end{center}
\caption{Feynman diagram singular in the collinear limit $p_m\parallel p_n$}
\label{rf3}
\end{figure}

 As mentioned before, the singularity occurs only if the collinear particles emerge from the same three gluon vertex, see Fig.3, therefore they appear in partial amplitudes in which $m$ is adjacent to $n$.
 Let us focus on such contribution to
$$2^{N/2}{\rm tr}(T^{a_1}\cdots T^{a_m}T^{a_n}\cdots T^{a_N} )A(1,\dots,m,n,\dots , N).$$
The propagator and three-gluon vertex contribute
\be \frac{-i}{2p_mp_n}(-i)(-g)\Big[(\epsilon_m\epsilon_n)(p_m-p_n)^\mu-2(\epsilon_np_m)
\epsilon_m^{\mu}+2(\epsilon_mp_n)\epsilon_n^{\mu}
\Big]\label{col3}\ee
If two helicities are identical, $h_n=h_m=h$, we can set  $\epsilon_m\epsilon_n=0$ by choosing the same reference spinor for both gluons. Then $\epsilon_m^{\mu}=\epsilon_n^{\mu}=\epsilon_P^{\mu}$ at the leading order and
\be\label{col7}A(1,\dots,m^h,n^h,\dots , N)\to \frac{g}{\sqrt 2} \frac{(\epsilon_mp_n-\epsilon_np_m)}{p_mp_n}A(1,\dots,P^h,\dots , N).\ee
This can be expressed in terms of momentum spinors as
\begin{eqnarray}A(1,\dots,m^-,n^-,\dots , N)&\to&  g \frac{1}{{\bf c\, s}[mn]}A(1,\dots,P^-,\dots , N)\\
A(1,\dots,m^+,n^+,\dots , N)&\to& g \frac{1}{{\bf c\, s}\langle mn\rangle}A(1,\dots,P^+,\dots , N).\end{eqnarray}\bs
{\bf Exercise 8.2} Derive the above result from Eq.(\ref{col7}) and check all factors of $\sqrt 2$ and signs.\es
\noindent Note that although the propagator pole is of order $1/\epsilon^2$, the collinear limit yields only a simple $1/\epsilon$ pole. This fact complicates discussion of the opposite helicity configuration, $h_n=-h_m=\pm 1$, because the first term in Eq.(\ref{col3}) cannot be eliminated by a smart choice of reference spinors. The best we can do is to choose the spinors $\lambda_r$ and $\tilde\lambda_r$ that appear in the definition of the collinear limit. Then $\epsilon_m\epsilon_n=-1$ and the double pole cancels as a consequence of gauge invariance. At the end, one finds
\begin{eqnarray}A(1,\dots,m^+,n^-,\dots , N)&\to&  g \frac{{\bf c}^3}{{\bf s}[mn]}A(1,\dots,P^+,\dots , N)\label{col4}\\ && +~ g\frac{{\bf s}^3}{{\bf c}\langle mn\rangle}A(1,\dots,P^-,\dots , N)\nonumber \end{eqnarray}\bs
{\bf Exercise 8.3} Starting from Eq.(\ref{col3}) and equations defining the collinear limit, derive  Eq.(\ref{col4}) and check all factors of $\sqrt 2$ and signs. Derive a similar expression for the collinear limit of $A(1,\dots,m^-,n^+,\dots , N)$.\es
\noindent Recently some sub-leading collinear terms of order  $\epsilon^0$ were also discussed \cite{Stieberger:2015kia}.
 \renewcommand{\refname}{\large Recommended Reading for Section 8}
  
\section{Factorization and BCFW Recursion}\setcounter{equation}{0}
\subsection{Factorization in Multi-Particle Channels}
The soft and collinear limits are two examples in which the propagator singularity can be extracted as a factor multiplying certain finite amplitude. In both cases, factorization occurs in a two-particle ``channel'' $(i,j)$ with $(p_i+p_j)^2\to 0$. In  general, singularities appear also in multi-particle channels at special values of the momenta $\sum_{i}p_i\equiv P$ when the combined momentum $P$ is tuned to $P^2=0$. Such singularity is due to Feynman diagrams with $P$ propagating along some internal (virtual) propagator lines, when the diagram can be separated into two parts connected by such a propagator. In order to understand factorization in multi-gluon channels let us focus on the partial amplitude $A(1,2,3,\dots, N)$ associated to the group trace $tr(T^{a_1}T^{a_2}T^{a_3}\cdots T^{a_N})$. If a Feynman diagram is separated into two pieces, called left and right,
connected by a propagator, the momentum flowing into the propagator must be a ``region'' momentum
$P_{i,m}\equiv p_i+p_{i+1}+p_{i+2}+\dots +p_{i+m-1}$ of the subset of $m$ momenta ordered in the same way as in the partial amplitude, see Fig.4.
\begin{figure}[h!]
\begin{center}
\includegraphics[scale=0.8]{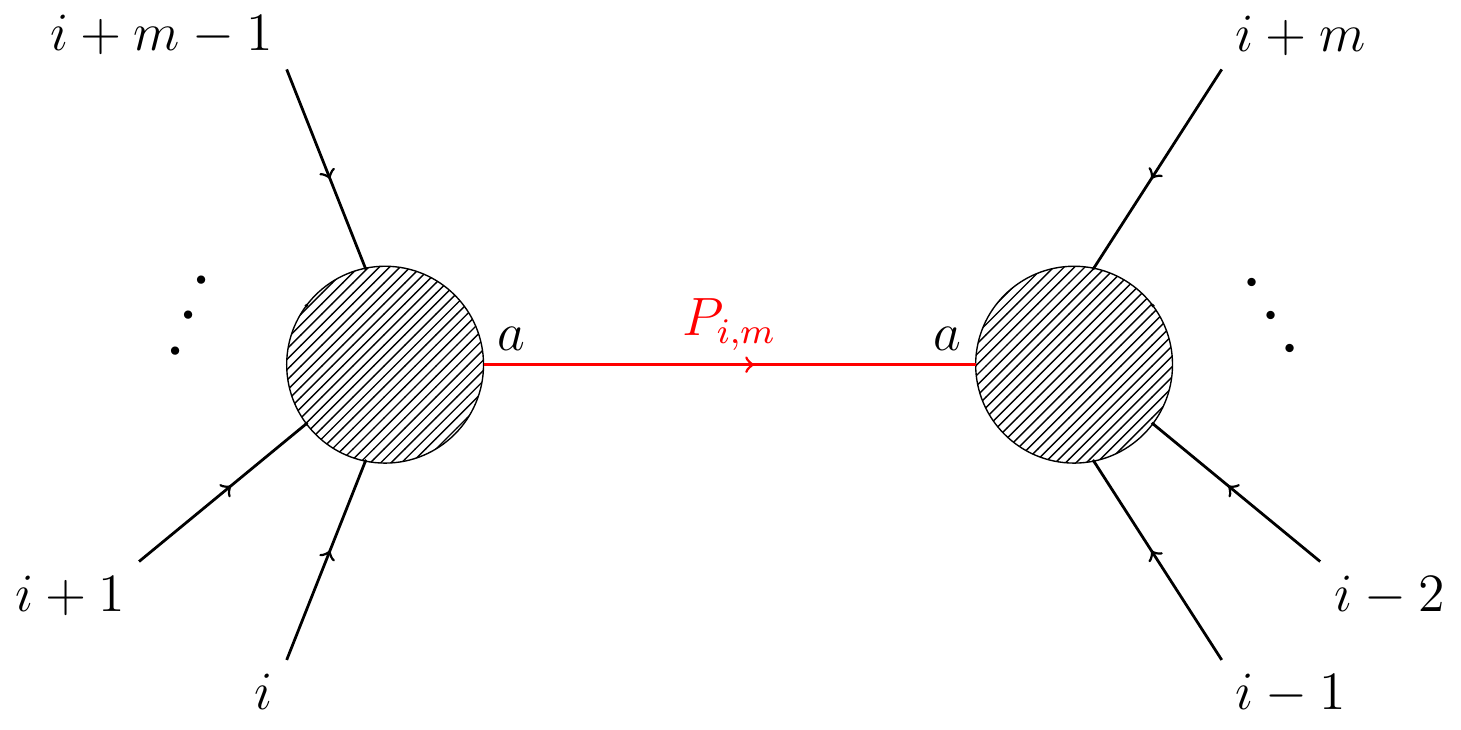}
\end{center}
\caption{Factorization in $m$-particle channel.}
\end{figure}
The reason is that the trace factor of the full diagram factorizes into the left and right parts connected as
\begin{eqnarray}\sum_a 2^{\frac{m+1}{2}}tr(T^{a_i}T^{a_{i+1}}\!\!&\cdots&\!\! T^{a_{i+m-1}}T^a)\, {2}^{\frac{N-m+1}{2}}tr(T^aT^{a_{i+m}}\cdots T^{a_N}T^{a_1}\cdots T^{a_{i-1}})\nonumber\\ && =~{2}^{\frac{N}{2}}\,tr(T^{a_1}T^{a_2}T^{a_3}\cdots T^{a_N})
\end{eqnarray}
therefore the momentum sum  must contain the same subset in order for $P_{i,m}$ to flow through the propagator. On the other hand, the kinematic part of the diagram splits as
\be (-i)A(i,i+1,\dots, i+m-1,\mu)\frac{-ig^{\mu\nu}}{P^2}(-i)A(\nu,i+m,\dots,N,1,\dots,i-1)\ ,\ee
where $P\equiv P_{i,m}$. Now we can replace
\be -g^{\mu\nu}=\sum_{h=\pm} \epsilon^{\mu *}_h(P,r)\epsilon^{\nu }_h(P,r)-\frac{P^\mu r^\nu+P^\nu r^\mu}{pr}\label{hss}\ee
and use gauge invariance to drop the second term. In this way, we obtain the following {\bf factorization theorem}: In the $P^{2}_{i,m}=P^2\to 0$ limit
\begin{eqnarray}&& A(1,2,3,\dots, N)~~\longrightarrow\label{factt}   \\ &&\sum_h
A(i,i+1,\dots, i+m-1,-P^{-h})\frac{1}{P^2}A(P^h,i+m,\dots,N,1,\dots,i-1)\ .\nonumber\end{eqnarray}
This theorem obviously holds for other partial amplitudes which factorize into partial subamplitudes involving the respective region momenta. Furthermore, it is also valid for amplitudes factorizing on intermediate fermion poles. This follows after replacing Eq.(\ref{hss}) by $P_{\alpha\dot\beta}=\lambda_\alpha(P)\bar\lambda_{\dot\beta}(P)$ on the internal fermion line.
\subsection{BCFW Recursion Relations}
 Factorization theorems allow recursive construction of $N$-particle amplitudes from the amplitudes involving smaller numbers of external particles. The idea is to ``deform'' one of momenta, say $p_i$, by a complex variable and tune it in order to hit a propagator pole and factorize full amplitude into simpler subamplitudes. While using complex momenta is perfectly OK, one needs to make sure that such a deformation does not violate the momentum conservation law, so actually at least two momenta have to be deformed, in the following way:
\be p_i^\mu\to \hat p_i^\mu(z)=p_i^\mu-zn^\mu,\qquad p_j^\mu\to \hat p_j^\mu(z)=p_j^\mu+zn^\mu\ ,\ee
where $z$ is a (adjustable) complex parameter and $n^\mu$ is a light-like vector satisfying
$$n^2= p_in= p_jn=0\ .$$
There are only two possible choices for such a vector:
\be n_{\alpha\dot\beta}=\lambda_{i\alpha}\tilde\lambda_{j\dot\beta}\ ~~{\bf (I)}\qquad{\rm or}\qquad
n_{\alpha\dot\beta}=\lambda_{j\alpha}\tilde\lambda_{i\dot\beta}\  ~~{\bf(II)} .\label{nch}\ee
The reason for such a choice of $n$ is that any channel with the region momentum sum involving $\hat p_i$ or $\hat p_j$ (but not both of them) can be ``put'' on-shell by a suitable choice of $z$:
\be \Big(\, \underbrace{\sum_{l\neq i,j} p_l+\hat p_i}_{\hat P_J(z)}\Big)^2=\Big(\,\underbrace{ \sum_{l\neq i,j} p_l+p_i}_{P_J}\Big)^2-2n\Big(\underbrace{ \sum_{l\neq i,j} p_l+p_i}_{P_J}\Big)z\label{zpole}\ee
The propagator singularity appears now as a simple pole in $z$. The deformed amplitude contains poles in all possible channels, where it factorizes into ``left'' and ``right'' parts with $\hat p_i$ and $\hat p_j$ always on the opposite sides as shown in Fig.5. At this point, complex analysis becomes helpful.
\begin{figure}[h!]
\begin{center}
\includegraphics[scale=0.8]{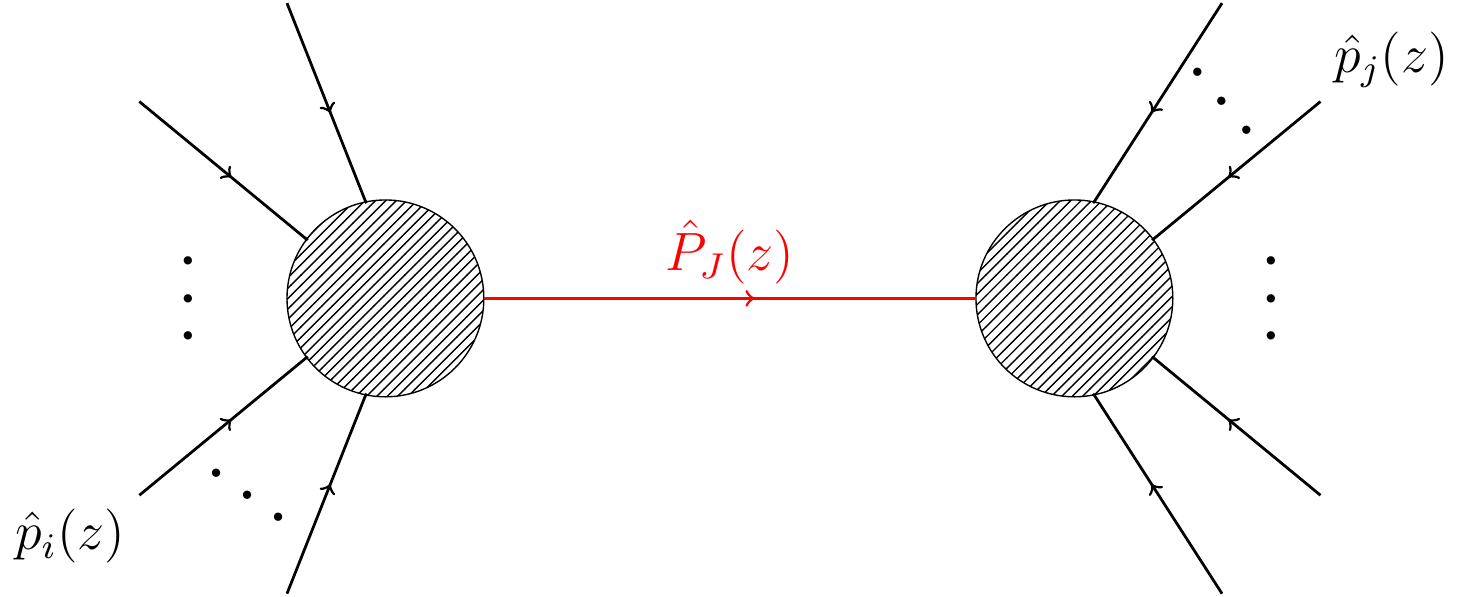}
\end{center}
\caption{Contribution to the residues from an on-shell propagator. For a $z$-dependent momentum flow
through the indicated internal propagator, particle $i$ has to be on the one side and particle $j$ has to
be on the other side.}
\label{fig_on_shell_residue}
\end{figure}

Consider a given deformed partial amplitude as a function of $z$:
\be A(z)=A\big(p_1^{h_1},\dots,\hat p_i^{h_i}(z),\dots,\hat p_j^{h_j}(z),\dots,p_N^{h_N}\big)\ee
Since tree amplitudes are rational functions of the momenta, $A(z)$ is a rational function of $z$,
variables. Let us assume that $A(z)$ falls off at least like $1/z$ for $|z|\to\infty$.
I will discuss the
detailed conditions for this to happen in the next paragraph. With this assumption we have
\be\frac{1}{2\pi i}\oint\frac{A(z)}{z}=0\ ,\ee
where the contour is a large circle at $|z| =\infty$ oriented counter-clockwise. On the other hand we
may evaluate this integral with the help of Cauchy's residue theorem. There is one residue at
$z = 0$ due to the explicit factor of $1/z$ in the integrand. This residue gives
\be A(0)=A\big(p_1^{h_1},\dots,p_i^{h_i},\dots,p_j^{h_j},\dots,p_N^{h_N}\big)\ee
which is the undeformed amplitude we want to calculate. All other residues come from internal
propagators of the amplitude. We have to consider only the propagators that are $z$-dependent.
A typical configuration is shown in Fig.5.
As mentioned before, particles $i$ an $j$ must be on opposite sides in order to have a $z$-dependent intermediate
propagator. Let us denote the
set of external legs on the one side by $I$ and the set of external legs on the other side by $J$.
Let us further assume that the multiplicity of these sets are $n_I=m$ and $n_J=N-m$, respectively.
We set
\bq
 P_J & = & -\sum\limits_{j \in J} p_j=\sum\limits_{i \in I} p_i
\eq
The momentum flowing through the internal propagator is
\bq
 \hat{P}_J(z) & = & P_J - z n.
\eq
According to Eq.(\ref{zpole}), the internal propagator goes on-shell for
\bq
 \hat{P}_J^2 \; = \; 0
 & \Rightarrow &
 z_J \; = \;
 \frac{P_J^2}{2 nP_J }.\label{zzpole}
\eq
In this limit the amplitude factorizes as in Eq.(\ref{factt})
and the residue is given by
\bq
 -
 \sum\limits_{h}
 A\Big(\dots, \hat{p}_i^{h_i}(z_J), \dots, -\hat{P}_J^{-h}(z_J)\Big)
 \frac{1}{P_J^2}
 A\left(\hat{P}_J^{h}(z_J),\dots, \hat{p}_j(z_J)^{h_j}, \dots\right).\nonumber
\eq
Summing over all residues we obtain the Britto-Cachazo-Feng-Witten (BCFW) {\bf on-shell recursion relation}:
\begin{eqnarray}
 A\Big(p_1^{h_1},\dots&&\!\!\!\!, p_i^{h_i}, \dots, p_j^{h_j}, \dots, p_N^{h_N}\Big)
 =  \label{bcfw}\\
 \sum\limits_{\mathrm{partitions}\; (I,J)}
 &&\!\!\!\!\!\!\sum\limits_{h}
 A\left(\dots, \hat{p}_i^{h_i}(z_J), \dots, -\hat{P}_J^{-h}(z_J)\right)
 \frac{1}{P_J^2}
 A\left(\hat{P}_J^{h}(z_J),\dots, \hat{p}_j(z_J)^{h_j}, \dots\right),
 \nonumber
\end{eqnarray}
where the sum is over all partitions $(I,J)$ such that particle $i \in I$ and particle $j \in J$.
Note that although the momentum $\hat{P}_J(z_J)$ is on-shell ($\hat{P}_J^2(z_J)=0$), the momentum $P_J$ appearing in the
denominator is in general not on-shell ($P_J^2 \neq 0$).
Eq.(\ref{bcfw}) allows us to compute the $N$-particle amplitude recursively through
on-shell amplitudes with fewer external legs.
In order to apply this recursion relation we have to ensure that the amplitude $A(z)$ vanishes
at $|z|\rightarrow \infty$.

Let us now consider the momenta shifts and the behaviour at $|z| = \infty$ in more detail.
In a given diagram, $z$ flows from $p_i$ to $p_j$ through a number of propagators and vertices.
The most dangerous contribution comes from a path where all vertices are three-gluon vertices that grow like $z$ for large $z$. The propagators are suppressed like $1/z$. For any path, each vertex is followed by a propagator, except for the last one,
therefore the product of vertices and propagators
behaves like $z$ for large $z$. What about external wave functions? In case ({\bf I}) of deformation by vector $$n_{\alpha\dot\beta}=\lambda_{i\alpha}\tilde\lambda_{j\dot\beta}$$  of Eq.(\ref{nch}), the momentum spinors change as
\bq
\label{shift}
 \hat \lambda_i=\lambda_i~~~~~~~~~~~~~~~~~~
 & &\hat{\tilde\lambda}_i=\tilde\lambda_i-z\tilde\lambda_j
 \nonumber \\ \hat{\lambda}_j=\lambda_j+z\lambda_i~~~~~~~~~\, & &\hat{\tilde\lambda}_j=\hat{\tilde\lambda}_i
\eq
Then the polarization vectors behave as
\bq
\label{pols}
 \epsilon_-(\hat p_i)\sim z^{-1}~~~~~~~~~~~~~~
 & &\epsilon_+(\hat p_i)\sim z
 \nonumber \\ \epsilon_-(\hat p_j)\sim z~~~~~~~~~~~~~~~~\, & &\epsilon_+(\hat p_j)\sim z^{-1}
\eq
It is clear that for $(h_i,h_j)=(-,+)$, this deformation gives rise to a sufficiently suppressed amplitude. A more subtle argument shows that it is also good for $(-,-)$ and $(+,+)$. {}For $(+,-)$, deformation ({\bf II}) does the job.

As an example, let us compute $A(1^-,2^+,3^-,4^+)$ by using BCFW recursion relation. Let's ``mark'' $(i,j)=(1,2)$,
which is a $(-,+)$ helicity configuration allowed by the shift (\ref{shift}). There is only one partition possible $(I=\{4,1\}, J=\{2,3\})$. According to (\ref{zzpole}), for $z_j=[14]/[24]$ the shifted momentum $\hat P_J=\hat p_1(z)+p_4$ goes on-shell:
$$\hat P_J=-\lambda_3\tilde\lambda_4\frac{[23]}{[24]}~~~\Rightarrow~ ~~ \lambda_J=\lambda_3,\quad\tilde\lambda_J=-\tilde\lambda_4\frac{[23]}{[24]}$$
On the other hand, the undeformed $P_J^2=2p_1p_4=2p_2p_3$ is {\em not} on shell. The amplitude factorizes into two 3-point amplitudes as in Eq.(\ref{bcfw}), but only $(h=+)$ contributes to the sum.\bs
{\bf Exercise 9.1} Explain why $(h=-)$ does not contribute and use Eq.(\ref{bcfw}) to show that
\be A(1^-,2^+,3^-,4^+)=g^2\frac{\langle 13\rangle^4}{\langle 12\rangle\langle 23\rangle\langle 34\rangle\langle 41\rangle}\label{mhv4}\ee
{\bf Exercise 9.2}. Use BCFW recursion to show that
\be A(1^+,2^+,3^+,4^+)=A(1^-,2^+,3^+,4^+)=0.\label{nmhv4}\ee
{\bf Exercise 9.3}. Use BCFW recursion to show by mathematical induction that
\be A(1^+,2^+,3^+,\dots, N^+)=A(1^-,2^+,3^+,\dots,N^+)=0.\label{nnmhv4}\ee\es
The amplitude (\ref{mhv4}) belongs to a sequence of ``mostly-plus'' $(--+++\dots)$ MHV ``maximally helicity violating'' amplitudes with all except two gluons carrying positive helicities. This sequence starts at $A(1^-,2^-,3^+)$ and continues to
\be A(1^+,2^+,\dots m^-,\dots, n^-,\dots, N^+)=g^{N-2}\frac{\langle mn\rangle^4}{\langle 12\rangle\langle 23\rangle\cdots\langle N1\rangle}\label{mhvn}\ee
Similarly, the ``mostly-minus'' MHV amplitudes are given by
\be A(1^-,2^-,\dots m^+,\dots, n^+,\dots, N^-)=g^{N-2}\frac{[mn]^4}{[12][23]\cdots[N1]}\ .\label{amhvn}\ee\bs
{\bf Exercise 9.4}. Use BCFW recursion to prove Eqs.(\ref{mhvn}) and (\ref{amhvn}) by mathematical induction.\es
\renewcommand{\refname}{\large Recommended Reading for Section 9}
  
\section{Supersymmetry Hiding at the Tree Level}\setcounter{equation}{0}
\subsection{Supersymmetry}
 Let us consider Yang-Mills theory coupled to a chiral fermion field $\chi$ in the adjoint representation, defined by the following Lagrangian:
\be\Lagr_{SYM} =-\frac{1}{4} F_{\mu\nu}^aF^{\mu\nu}_a + i\bar{\chi}^a\bar\sigma^\mu D_\mu\chi_a +\frac{1}{2} D^aD_a\ ,
\label{sym}\ee
where the last term involves an additional ``auxiliary'' scalar field $D^a$ introduced for some technical reasons that will be explained later.  Recall that
$$ F_{\mu\nu}^a=\partial_\mu A^a_\nu-\partial_\nu A^a_\mu-gf_{abc}A_\mu^bA_\nu^c.$$
{}For a field transforming in the adjoint representation, the covariant derivative reads
$$D_\mu \chi^a=\partial_\mu \chi^a-gf_{abc}A_\mu^b\chi^c.$$
Since fermions transform in the same representation as gauge bosons, one expects some type of boson-fermion symmetry. Indeed, there is a group of transformations labeled by a constant chiral spinor parameter $\eta_\alpha$ that leaves the Lagrangian (\ref{sym}) invariant:
\begin{eqnarray}
\delta_\eta A^{a\mu} &=& -i\bar\chi^a\bar\sigma^\mu\eta+i\bar\eta\bar\sigma^\mu\chi^a\nonumber\\
\delta_\eta \chi^a &=&\frac{1}{2}F_{\mu\nu}^a\sigma^\mu\bar\sigma^\nu\eta +i\eta D^a\label{transf}\\[1mm]
\delta_\eta D^a &=&\eta\sigma^\mu D_\mu\bar\chi^a+D_\mu\chi^a\sigma^\mu\bar\eta\nonumber
\end{eqnarray}\bs
{\bf Exercise 10.1} Show that the Lagrangian (\ref{sym}) is indeed invariant under the above transformations.\es
\noindent This symmetry is called ``supersymmetry'' because unlike the case of ordinary symmetries, the transformation parameter is a spinor. This means that the corresponding symmetry charge (supercharge) is also a spinor ($Q_{\alpha},\bar Q^{\dot\alpha}$), so that for any field $\phi$
\be\delta_\eta \phi=[iQ(\eta),\phi]~~~{\rm with}~~~ Q(\eta)\equiv\eta^\alpha Q_\alpha +\bar\eta_{
\dot\alpha} \bar Q^{\dot\alpha}\label{qtr}\ee
Since $Q_\alpha$ and $\bar Q^{\dot\alpha}$ are spinor charges, they should obey certain anti-commutation relations, with the algebra closing on a set of symmetry operators. Furthermore, the transformation spinors $(\eta_1,\eta_2)$ should also be regarded as anti-commuting (Grassmann) variables with the property $\eta_1\eta_2= -\eta_2\eta_1$, $(\eta_1)^2=(\eta_2)^2=0$.\bs
{\bf Exercise 10.2} Show that the transformations (\ref{transf}) are consistent with the supersymmetry algebra
\be \{Q_\alpha,\bar Q_{\dot\alpha}\}=-2i\sigma^\mu_{\alpha\dot\alpha}\partial_\mu\label{mom}\ee
\be \{Q_\alpha, Q_{\beta}\}=\{\bar Q_{\dot\alpha},\bar Q_{\dot\beta}\}=0\ .\ee\es
\noindent Note that the anticommutator (\ref{mom}) closes on the familiar momentum operator, which belongs to the Poincare symmetry group. The theory defined by (\ref{sym}) is called super-Yang-Mills (SYM) theory and the fermion $\chi$ is called ``gaugino''. It is a chiral fermion transforming in the real representation of the gauge group, therefore its own antiparticle. Gaugino is an example of a ``Majorana'' fermion. The gaugino and gauge boson form a vector
``supermultiplet'' of supersymmetry.

In this course, not much attention is paid to scalar particles although they enter in many extensions of SYM, in particular in ``maximally supersymmetric'' ${\cal N}=4$ supersymmetric Yang-Mills theory. This theory is interesting
because it has a superconformal symmetry surviving all orders of perturbation theory. Together with Weyl fermions, scalars enter into ``chiral'' multiplets. Actually, the simplest example of a supersymmetric theory is the free theory of one complex scalar field $\phi$ and one left-handed fermion $\psi$ described by the Lagrangian
\be\Lagr_{WZf} =\partial^\mu\phi^*\partial_\mu\phi+ i\bar{\psi}\bar\sigma^\mu \partial_\mu\psi +F^*F\ ,
\label{wz}\ee
which is (super)symmetric under
\begin{eqnarray}
\delta_\eta\phi &=& -\sqrt 2 \eta\psi\nonumber\\
\delta_\eta \psi &=&i\sqrt 2\sigma^\mu\bar\eta\partial_\mu\phi-\sqrt 2\eta F\\
\delta_\eta F &=&i\sqrt 2\bar\eta\bar\sigma^\mu\partial_\mu\psi\nonumber
\end{eqnarray}
Here, $F$ is another auxiliary field; $F=0$ after using field equations. The Lagrangian (\ref{wz}) can be generalized to include supersymmetric interactions of fermions and scalars, as well as their interactions with  vector supermultiplets.
For example, supersymmetric extensions of the standard model involve quarks and squarks interacting with gauge bosons and gauginos. On the other hand, ${\cal N}=4$ SYM can be constructed by coupling three chiral multiplets in the adjoint representation to the gauge vector multiplet.
\subsection{Supersymmetric Ward Identities}
 In  SYM, the amplitudes involving gauginos and gauge bosons must be somehow related. Since the amplitudes are defined as vacuum expectation values involving in/out creation/annihilation operators creating free particles in the asymptotic regions, the first task is to understand how supersymmetry acts on these operators. The free fields are
\be A^{\mu}(x)=\int\frac{d^3\pb}{(2\pi)^3}\frac{1}{\sqrt{2E_p}}\,\sum_{h=\pm}\Big[\epsilon_h^{\mu}(p,r)e^{-ipx}
a_{\mathbf{p}}^h+\epsilon_{h}^{*\mu}(p,r)e^{ipx}
a_{\mathbf{p}}^{h\dag}\Big]\ee
\be\chi_\alpha(x) = \int \frac{d^3\pb}{(2\pi)^3}\frac{1}{\sqrt{2E_p}}\,\lambda_{\alpha}({\pb})\Big[
b_{\pb}^-e^{-ipx}+
b_{\pb}^{+\dagger}e^{ipx}\Big]\, \ee
{}For free fields, the supersymmetry transformations (\ref{transf}) read
\begin{eqnarray}
[iQ(\eta), A^a_\mu] &=& -i\bar\chi^a\bar\sigma^\mu\eta+i\bar\eta\bar\sigma^\mu\chi^a\nonumber\\[1mm]
[iQ(\eta), \chi^a ]&=&\frac{1}{2}(\partial_\mu A_\nu^a-\partial_\nu A_\mu^a)\sigma^\mu\bar\sigma^\nu\eta
\end{eqnarray}
where we set $g=0$ and used free equations of motion, including $D^a=0$.\footnote{$D^a$ was introduced in the first place for the sole purpose of closing supersymmetry algebra off-shell, i.e.\ without using equations of motion.} When free fields are inserted into the above transformations, the l.h.s.\ brings commutators of $Q(\eta)$ with the creation and annihilation operators and  can be compared to the r.h.s. To that end, it is convenient to choose the reference spinor $r=\eta$ for the polarization vectors. For annihilation operators, we get
\begin{eqnarray}
[iQ(\eta),a^-_{\mathbf{p}}]=i\sqrt 2\,[\bar \lambda({\pb})\bar\eta]\, b^-_{\mathbf{p}} \quad&&\quad [iQ(\eta),a^+_{\mathbf{p}}]=-i\sqrt 2\,\langle \lambda({\pb})\eta\rangle\, b^+_{\mathbf{p}}\label{qcom}\\[2mm]
[iQ(\eta),b^-_{\mathbf{p}}]=-i\sqrt 2\,\langle \lambda({\pb})\eta\rangle\, a^-_{\mathbf{p}}~ \quad&&\quad
[iQ(\eta),b^+_{\mathbf{p}}]=i\sqrt 2\,[\bar \lambda({\pb})\bar\eta]\, a^+_{\mathbf{p}}\nonumber
\end{eqnarray}
while the commutators with creation operators are obtained by hermitean conjugation.\bs
{\bf Exercise 10.3} Derive Eqs.(\ref{qcom}) and check all signs, factors of  $i$ and $\sqrt 2$.\es
The free fields of the chiral multiplet are:
\be \phi(x)=\int\frac{d^3\pb}{(2\pi)^3}\frac{1}{\sqrt{2E_p}}\Big[s_{\mathbf{p}}^-e^{-ipx}
+s_{\mathbf{p}}^{+\dagger} e^{ipx}\Big]\label{ssf}\ee
\be\psi_\alpha(x) = \int \frac{d^3\pb}{(2\pi)^3}\frac{1}{\sqrt{2E_p}}\,\lambda_{\alpha}({\pb})\Big[
d_{\pb}^-e^{-ipx}+
d_{\pb}^{+\dagger}e^{ipx}\Big]\, \label{freec}\ee
Now the relevant commutators are
\begin{eqnarray}
[iQ(\eta),d^-_{\mathbf{p}}]=\sqrt 2\,[\bar \lambda({\pb})\bar\eta]\, s^-_{\mathbf{p}} \quad&&\quad [iQ(\eta),d^+_{\mathbf{p}}]=\sqrt 2\,\langle \lambda({\pb})\eta\rangle\, s^+_{\mathbf{p}}\label{qcoms}\\[2mm]
[iQ(\eta),s^-_{\mathbf{p}}]=\sqrt 2\,\langle \lambda({\pb})\eta\rangle\, d^-_{\mathbf{p}}~ \quad&&\quad
[iQ(\eta),s^+_{\mathbf{p}}]=\sqrt 2\,[\bar \lambda({\pb})\bar\eta]\, d^+_{\mathbf{p}}\nonumber
\end{eqnarray}
From now on, we will absorb the factors $\sqrt 2$ in Eqs.(\ref{qcom}) and (\ref{qcoms}) into the definition of $\eta$.

The relations between scattering amplitudes implied by symmetries of the theory are called Ward identities and can be obtained in the following way. Since the vacuum is invariant under symmetry transformations,
$$e^{iQ}|0\rangle= |0\rangle~~~\Rightarrow~~~~{Q}|0\rangle=0\ ,$$
i.e.\ the vacuum is annihilated by the symmetry generators. Consider the vacuum expectation value of the following  chain of in/creation and out/annihilation operators:
$\langle 0|a_1a_2\cdots a_N|0\rangle$.
If $Q$ is a symmetry generator, then
$\langle 0|[Q,a_1a_2\cdots a_N]|0\rangle=0$, which after applying Leibniz chain rule gives
\be\langle 0|[Q,a_1]a_2\cdots a_N]  |0\rangle +  \langle 0|a_1[Q,a_2]\cdots a_N]  |0\rangle
+\dots + \langle 0|a_1a_2\cdots [Q,a_N]  |0\rangle=0\ .\ee
 This is an example of Ward identity which yields relations among amplitudes involving creation/ annihilation operators related by symmetry transformations. Note that since supersymmetry does not act on gauge indices, SYM Ward identities hold at the level of partial amplitudes.

 As an example, consider a four-gluon amplitude $A(1^-2^+3^-4^+)$ which originates from the following correlator:
 \be \langle 0|a_1^+a_2^-a_3^+a_4^-|0\rangle\ee
 where we flipped helicities in order to take into account our convention that all particles are incoming.
 We can derive the following Ward identity
 \begin{eqnarray}&&[1\bar\eta]\langle 0|a_1^+a_2^-a_3^+a_4^-|0\rangle=\langle 0|[Q(\eta),b_1^+]a_2^-a_3^+a_4^-|0\rangle\\[2mm] &=&[2\bar\eta] \langle 0|b_1^+b_2^-a_3^+a_4^-|0\rangle-
\langle 3\eta\rangle \langle 0|b_1^+a_2^-b_3^+a_4^-|0\rangle+
[ 4\bar\eta] \langle 0|b_1^+a_2^-a_3^+b_4^-|0\rangle\nonumber
\end{eqnarray}
Note that when pulling out $\eta$-dependent factors, we used the fact that $\eta$ anticommutes with fermionic $b$ operators.
From here, we obtain 3 relations. There is only one $\eta$ angle factor, in the middle term of the second line, therefore the coefficient must vanish:
\be A(1_\chi^-2^+3_\chi^-4^+)=0\ .\label{w1}\ee
This identity is obviously true because it also follows from the $U(1)$ ``$R$ symmetry'' of SYM which transforms $\chi\to e^{i\alpha}\chi$.
On the other hand, by choosing $\bar\eta=1$, we obtain
\be A(1_\chi^-2_\chi^+3^-4^+)= -\frac{[14]}{[12]}A(1_\chi^-2^+3^-4_\chi^+)\ \label{w2},\ee
and finally, by choosing $\bar\eta=4$, we obtain
\be A(1^-2^+3^-4^+)=\frac{[24]}{[14]}A(1_\chi^-2_\chi^+3^-4^+)\ \label{w3},\ee
which is equivalent to
\be A(1_\chi^-2_\chi^+3^-4^+)=-\frac{\langle 23\rangle}{\langle 13\rangle} A(1^-2^+3^-4^+) \label{w5}.\ee


It is quite possible that supersymmetry is a symmetry of nature, but this is not a lecture about physics beyond the standard model, so why do we bother with SUSY? The Feynman rules for SYM are essentially the same as for QCD with one massless, chiral (say left-handed) quark transforming in the adjoint instead of the fundamental representation. The basic coupling is the gaugino-gluon coupling involving two fermion lines emerging from the vertex. Let us consider pure $N$-gluon amplitudes, without any external fermion line. Is QCD different from SYM? There are no external fermion lines, but are there any internal lines possible? Since fermions are always created in pairs, they would have to end up in external lines or they would have to close the loop. Thus at the tree level, there is no difference between QCD, SYM or even pure YM! So we can rename our tree-level $N$-gluon amplitudes as SYM amplitudes and use Ward identites as a computational tool. For example Eq.(\ref{w3}) offers a shortcut to the four-gluon amplitude, by computing a simpler amplitude with two gauginos and two gluons. One could continue even farther and trade all gluons for gauginos. So even if SUSY is long forgotten as an extension of the standard model, it can always be used as a computational tool!\bs
{\bf Exercise 10.4}. Check Eq.(\ref{w2}) by computing the amplitudes on both sides.\\
{\bf Exercise 10.5}. Derive a SUSY Ward identity that allows expressing $A(1^-2^-3^+4^+)$ in terms of a  purely fermionic amplitude and compute it in order to check that it yields the correct four-gluon amplitude.\\
{\bf Problem 10.6}. Use SUSY Ward identities to prove that
\be A(1^+,2^+,3^+,\dots, N^+)=A(1^-,2^+,3^+,\dots,N^+)=0\ .\ee
{\bf Exercise 10.7.} ${\cal N}=4$ SYM is invariant under four supersymmetry transformations and contains four species of gauginos $(h=\pm 1/2)$ and three complex scalars $(h=\pm 0)$ in addition to the gauge boson. Each SUSY transformation pairs one particular gaugino with the gauge boson into a vector multiplet $(a,b)$, while pairing the remaining gauginos with scalars in three chiral multiplets $(d,s)_i, ~i=1,2,3$. By using SUSY Ward identities associated to one particular SUSY transformation, one can descend from gauginos to their scalar partners by using Eqs.(\ref{qcoms}). Show that
\be A(1^{-h},2^h,3^-,4^+)=D(h)\left(\frac{\langle 23\rangle}{\langle 13\rangle}\right)^{2(1-h)}A(1^-, 2^+ ,3^- ,4^+)\ .\label{ddh1}\ee \noindent where \be D\{-1,-1/2,0,1/2,1\}=\{1,1,1,-1,1\}\ .\label{ddh}\ee
\es
\renewcommand{\refname}{\large Recommended Reading for Section 10}

\section{Cutting Through One Loop}\setcounter{equation}{0}
\subsection{Regularization}
At the loop level, one encounters integrals over loop momenta, with the integrands containing denominators coming from the propagators and possibly numerator factors coming from both propagators and vertices. We will discuss the loop integrals with trivial numerators first. At one loop, all such ``scalar'' integrals can be reduced  to
\be \int\frac{d^4k}{(k^2-m^2+i0)^a}\label{tad}\ee
where $a$ are positive integers.
Naive power counting tells us that for $a=1$ such integral diverges quadratically at $k\to\infty$, while for
$a=2$ it contains a logarithmic divergence. These ``ultraviolet'' divergences need to be ``regularized'' in order to make sense out of loop corrections. The standard way of dealing with such divergences is by using dimensional regularization, by defining the theory in $D=4-2\epsilon$ dimensions with $\epsilon >0$.
 Thus the basic integral is
\be I(a,m^2)=\int\frac{d^Dk}{(-k^2+m^2-i0)^a}=
\int d^{D-1}\mathbf{k}\int dk_0\frac{1}{(-k_0^2+\mathbf{k}^2+m^2-i0)^a}
\label{tad1}\ee
Consider the integration over $k_0$. We see that there are two poles in the complex plane, at $k_0=
\pm\sqrt{\mathbf{k}^2+m^2}\mp i0$. We can (Wick-)rotate the contour of integration for $k_0$ in the complex plane from the real axis to the imaginary axis, $k_0=ik_{0E}$, without encountering these poles. Defining an Euclidean $D$-vector by $k_E=(k_{0E},\mathbf{k})$, we arrive at
\be i\int\frac{d^Dk_E}{(k^2_E+m^2)^a}\ee
In order to perform integrations, it is convenient to write the denominator in the Schwinger parametrization:
\be\frac{1}{(k^2_E+m^2)^a}=\frac{1}{\Gamma(a)}\int_0^\infty \,x^{a-1}e^{-x(k^2_E+m^2)}dx\ee
where $\Gamma$ is the Euler gamma function.

Let us recall some basic properties of this function. It is defined as
$$\Gamma(z)=\int_0^\infty x^{z-1}e^{-x}dx\ .$$
Note that for positive integers $n$, $\Gamma(n)=(n-1)!$. $\Gamma(z)$ has simple poles at $z=0$ and at $z$ equal to negative integers. In particular, near $z=0$,
$$\Gamma(z)=\frac{1}{z}-\gamma+\frac{1}{2}\Big(\gamma^2+\frac{\pi^2}{6}\Big)z+\dots$$
where $\gamma\approx 0.577\dots$ is the Euler-Mascheroni constant. The expansions near other poles can be obtained by using $z\Gamma(z)=\Gamma(z+1)$.

Now the momentum integration boils down to the Gaussian integral in $D$ dimensions:
$$\int d^Dk_E\, e^{-xk^2_E}=(\pi/x)^{D/2}.$$
We obtain
\be I(a,m^2)=i\pi^{2-\epsilon}\,\frac{\Gamma(a-2+\epsilon)}{\Gamma(a)}
(m^2)^{2-a-\epsilon}
\ee
The dimensionally-regularized integrals (\ref{tad}) are finite for all integer $a$, including the previously divergent cases of $a=1$ and $a=2$. These divergences appear in the $\epsilon\to 0$ limit as $1/\epsilon$ poles.

A generic one-loop Feynman diagram is depicted in Fig.6.
\begin{figure}[h!]
\begin{center}
\includegraphics[scale=1.6]{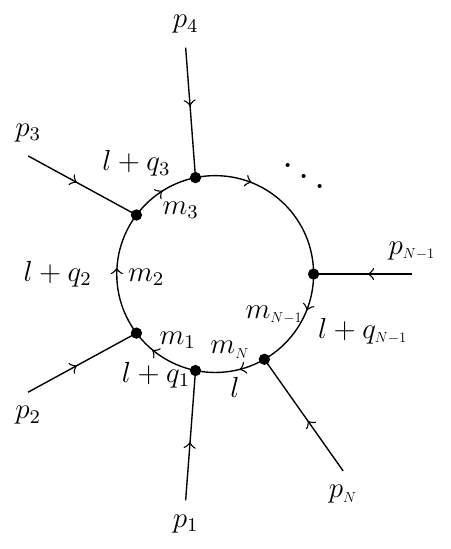}
\end{center}
\caption{Generic one-loop diagram. Here $q_j=\sum_{k=1}^jp_k$ are the ``region'' momenta.}
\end{figure}
 The diagram with just one external leg (and one internal propagator) is called the tadpole diagram. The tadpole integral is
\be I_1(m^2)=\int\frac{d^Dl}{(2\pi)^D}\frac{1}{l^2-m^2}=-(2\pi)^{-D}I(1,m^2),\ee
where
$$I(1,m^2)=i\pi^{2-\epsilon}\,{\Gamma(-1+\epsilon)}
(m^2)^{1-\epsilon}=i\pi^{2-\epsilon}\,\frac{\Gamma(\epsilon)}{(-1+\epsilon)}
(m^2)^{1-\epsilon}$$
The mass-dependence boils down to the factor
$$(m^2)^{1-\epsilon}=m^2(1-\epsilon\ln m^2+\dots),$$
  This integral appears in $\lambda\phi^4$ theory and in many other places. In $\lambda\phi^4$ theory with finite mass, the $m^2\epsilon^{-1}$ pole term can be removed by introducing a mass counterterm so
the net effect of tadpole correction is a logarithmic mass renormalization. Notice that in the limit $m\to 0$ the tadpole integral vanishes, which is a curious property of dimensional regularization.

The diagram with two external legs involves the bubble integral
\be I_2(p^2,m_1^2,m_2^2)=\int\frac{d^Dl}{(2\pi)^D}\frac{1}{[(l+p)^2-m_1^2](l^2-m_2^2)}\ee
In order to express it in terms of $I(a,\dots)$, the denominators can be combined by using the Feynman's parameter method:
\be\frac{1}{X_1^{a_1}X_2^{a_2}}=\frac{\Gamma(a_1+a_2)}{\Gamma(a_1)\Gamma(a_2)}\int_0^1d\alpha
\frac{\alpha^{a_1-1}(1-\alpha)^{a_2-1}}{[\alpha X_1+(1-\alpha)X_2]^{a_1+a_2}}\ee
Let's consider the simpler case of $m_1=m_2=m$. Then
\begin{eqnarray} \frac{1}{[(l+p)^2-m^2](l^2-m^2)}&=&\int_0^1d\alpha \frac{1}{(l^2+2\alpha lp +\alpha p^2-m^2)^2}\nonumber\\ &=&
\int_0^1d\alpha \frac{1}{[(l+\alpha p)^2 +\alpha(1-\alpha) p^2-m^2]^2}
\end{eqnarray}
After shifting the integration momentum to $k=l+\alpha p$, we obtain
\be I_2(p^2,m^2,m^2)=(2\pi)^{-D}\int_0^1I[2,m^2-\alpha(1-\alpha) p^2]\, d\alpha\label{dis1} \ee
with
\be\int_0^1I[2,m^2-\alpha(1-\alpha) p^2]\, d\alpha=i\pi^{2-\epsilon}\Gamma(\epsilon)\int_0^1[m^2-\alpha(1-\alpha) p^2]^{-\epsilon}d\alpha\ee
Note that the term in square bracket vanishes for
$$\alpha_{\pm}=\frac{1}{2}\pm \frac{\sqrt{1-{4m^2\over p^2}}}{2}$$
which give rise to two branch point singularities of the integrand.
\subsection{Discontinuities}
If $p^2<4m^2$, the integral (\ref{dis1}) can be easily performed without encountering branch points in the integration region.\bs
{\bf Exercise 11.1}. For $p^2< 4m^2$, expand $I_2(p^2,m^2,m^2)$ up to the order ${\cal O}(\epsilon^0)$ and perform the integral over Feynman parameter $\alpha$. If you are a Mathematica user, you can try to obtain an exact expression valid to all orders in $\epsilon$. You should obtain some type of a hypergeometric function.\es
\noindent For $p^2\geq 4m^2$, the Feynman parameter integral is ambiguous because it depends on  the choice of integration path, above or below the branch cut $(\alpha_- ,\alpha_+)$. The two choices differ by a finite number. One way of characterizing this ambiguity is by analytically continuing to complex $p^2$ and computing the discontinuity across the real axis:
\begin{eqnarray}{\rm Disc}_{p^2}I_2(p^2,m^2,m^2)&\equiv&\lim_{\eta\to 0}[I_2(p^2+i\eta,m^2,m^2)-I_2(p^2-i\eta,m^2,m^2)]\nonumber\\
&=&\frac{i\pi^2}{(2\pi)^4}\int_{\alpha_-}^{\alpha_+}(-2\pi i) d\alpha=\frac{1}{8\pi}\sqrt{1-{4m^2\over p^2}}\ \label{dis}\end{eqnarray}\bs
{\bf Exercise 11.2}. Show that for two particles with identical mass $m$ and total momentum $p=k_1+k_2$ ($E_i=\sqrt{\mathbf{k}_i^2+m^2}$), the volume of the Lorentz invariant phase space
\begin{eqnarray}
 \int dLIPS_2(k_1+k_2=p)&=&\int\frac{d^3\mathbf{k}_1}{(2 \pi)^3 2E_1}\int \frac{d^3\mathbf{k}_2}{(2 \pi)^3 2 E_2}(2\pi)^4 \delta^4(p-k_1-k_2)\nonumber\\
&=&\frac{1}{8\pi}\sqrt{1-{4m^2\over p^2}}\ .\label{lip}
\end{eqnarray}\es
It is not an accident that the discontinuity (\ref{dis}) is equal to the volume of the phase space (\ref{lip}).
In order to explain this point, it is good to place our computation in a specific context. In the present case it is a theory of one real scalar field $H$ with mass $M$ and one complex scalar field $\Phi$ with mass $m$, coupled to each other via
$${\cal L}_{\rm int}=\lambda H\Phi\Phi^*$$
where $\lambda$ is a dimension 1 coupling constant. The $H$ self-energy  correction is given by a bubble diagram with $\Phi$ propagating in the loop as shown in Fig.7. 
\begin{figure}[h!]
\begin{center}
\includegraphics[scale=1.0]{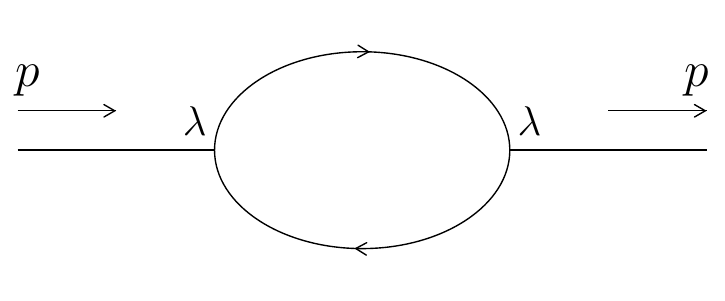}
\end{center}
\caption{One-loop self-energy correction.}
\end{figure}
It is
\be \Sigma_h(p^2)=\lambda^2I_2(p^2,m^2,m^2)\ .\ee
Note that if $p^2=M^2\geq 4m^2$, $H$ can decay into a $\Phi\bar\Phi$ particle-antiparticle pair. It is exactly in this region where $\Sigma_H(p^2)$ develops a discontinuity:
\be{\rm Disc}_{p^2}\Sigma_H(p^2) =\frac{\lambda^2}{8\pi}\sqrt{1-{4m^2\over p^2}}\label{sdis}\ee
According to the unitarity (Cutkosky) rules, if the loop can be cut into two parts in such a way that the disconnected sides of the Feynman diagram describe kinematically allowed processes (in this case $H$ decay into a $\Phi\bar\Phi$ pair and their recombination), the diagram develops a discontinuity determined by the product of respective amplitudes:
\be {\rm Disc}_{p^2}\Sigma_H(p^2) =-\sum\int dLIPS_2(k_1+k_2=p)\,A(\mathbf{p};\mathbf{k}_1,\mathbf{k}_2)\times\! A(\mathbf{k}_1,\mathbf{k}_2;\mathbf{p}).\ee
where the sum runs over all possible quantum numbers of intermediate particles. Here it is the $\Phi\bar\Phi$ pair. A graphical representation of this dicontinuity is shown in Fig.8.
\begin{figure}[h!]
\begin{center}
\includegraphics[scale=0.8]{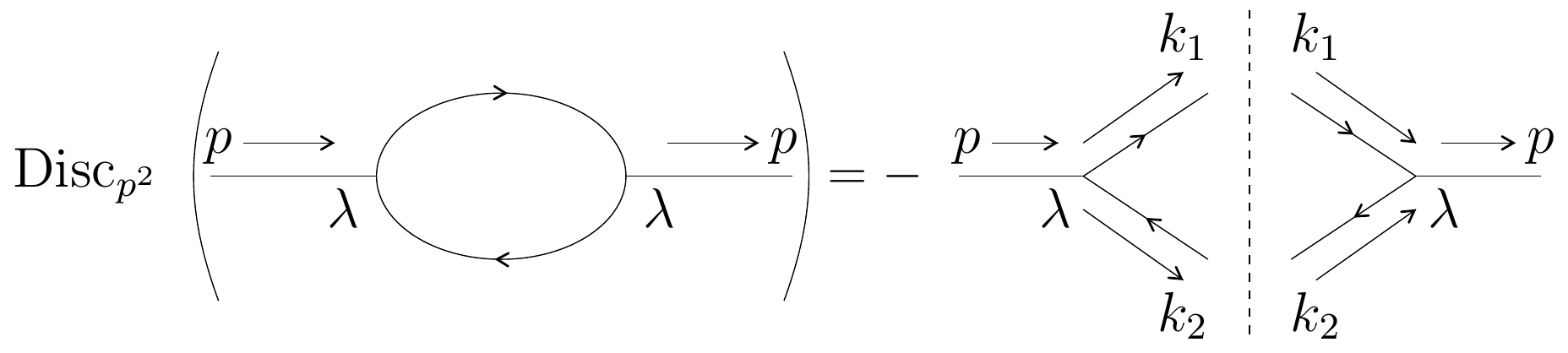}
\end{center}
\caption{Discontinuity of the self-energy correction.}
\end{figure}
Indeed, since $A(\mathbf{p};\mathbf{k}_1,\mathbf{k}_2)=A(\mathbf{k}_1,\mathbf{k}_2;\mathbf{p})=i\lambda$, we recover Eq.(\ref{sdis}).

As another example, let us apply Cutkosky rules to a theory with $H$ coupled to a  Dirac field $\Psi$ of mass $m$,
with dimensionless Yukawa coupling $y$:
\be {\cal L}_{\rm int}=y H\bar{\Psi}\Psi .\label{lyh}\ee
Now
\begin{eqnarray} \Sigma_H(p^2)&=&y^2 \int\frac{d^Dl}{(2\pi)^D}\frac{-{\rm Tr}[(\slashed{l}+\slashed{p}+m)(\slashed{l}+m)}{[(l+p)^2-m^2](l^2-m^2)}\nonumber\\
&=& -4y^2\int\frac{d^Dl}{(2\pi)^D}\frac{l^2+m^2+lp}{[(l+p)^2-m^2](l^2-m^2)}\label{sid}\\[2mm] &=&-4y^2\big[I_1(m^2)+2m^2I_2(p^2,m^2,m^2)+p_\mu I_2^\mu(p^2,m^2,m^2)\big],\nonumber
\end{eqnarray}
where
\be I_2^\mu(p^2,m^2,m^2)=\int\frac{d^Dl}{(2\pi)^D}\frac{l^\mu}{[(l+p)^2-m^2](l^2-m^2)}\label{teni}\ee
By using Feynman parametrization, we obtain
\be I_2^\mu(p^2,m^2,m^2)=(2\pi)^{-D}p^\mu\int_0^1I[2,m^2-\alpha(1-\alpha) p^2](-\alpha) d\alpha \ee\bs
{\bf Exercise 11.3} Prove the above result, expand it up to the order ${\cal O}(\epsilon^0)$ and perform the Feynman parameter integration. Mathematica users should also try to obtain an exact expression  for $I_2^\mu(p^2,m^2,m^2)$, valid to all orders in $\epsilon$.\\
{\bf Exercise 11.4} Show that
\be {\rm Disc}_{p^2}[p_\mu I_2^\mu(p^2,m^2,m^2)]=-\frac{p^2}{2}\left(\frac{1}{8\pi}\right)\sqrt{1-{4m^2\over p^2}}\ \ .\ee\es
\noindent The tadpole integral $I_1$ has no discontinuity because does not depend on $p^2$. Hence
\be {\rm Disc}_{p^2}\Sigma_H(p^2)=y^2(2p^2-8m^2)\left(\frac{1}{8\pi}\right)\sqrt{1-{4m^2\over p^2}}\label{sdd}\ee
In order to compare the above result with the discontinuity obtained by using Cutkosky rules, we compute
\begin{eqnarray} -\sum  \int && \!\!\!\!\!\! dLIPS_2\, A(\mathbf{p};\mathbf{k}_1,\mathbf{k}_2)\times\! A(\mathbf{k}_1,\mathbf{k}_2;\mathbf{p})\nonumber\\ && =y^2\sum_{s,s'}\int dLIPS_2 \,[\bar u^s(\mathbf{k}_1)v^{s'}(\mathbf{k}_2)]\times [\bar v^{s'}(\mathbf{k}_2)u^{s}(\mathbf{k}_1)]\\
&& =y^2\int dLIPS_2 \,{\rm Tr}[(\slashed{k}_1+m)(\slashed{k}_2-m)]\ ,\nonumber
\end{eqnarray}
where we used Eqs.(\ref{com1}) and (\ref{com2}).
After using
$${\rm Tr}[(\slashed{k}_1+m)(\slashed{k}_2-m)]=4(k_1k_2-m^2)=2p^2-8m^2\ ,$$
we recover the discontinuity (\ref{sdd}). It is clear that Cutkosky rules determine discontinuities very efficiently, making use of on-shell amplitudes instead of complicated loop integrals. To what extent the loop diagrams can be determined by (or constructed from) the on-shell amplitudes?
\renewcommand{\refname}{\large Recommended Reading for Section 11}
  
\section{Elements of Unitarity Cut Method}\setcounter{equation}{0}
\subsection{Reconstructing from a cut}
In the previous Section, we applied Cutkosky rules to a theory with the scalar field $H$ coupled to a  Dirac field $\Psi$ of mass $m$,
with dimensionless Yukawa coupling $y$, c.f.\
 Eq.(\ref{lyh}).
In this context, we evaluated the ``vector'' integral
$I_2^\mu(p^2,m^2,m^2)$, defined in Eq.(\ref{teni}),
by using Feynman parametrization etc. Actually, it was the most inefficient way of treating this integral. Here is a better way to go. We know that $I_2^\mu(p^2,m^2,m^2)\sim p^\mu$, so we know that
\be I_2^\mu(p^2,m^2,m^2)=c_1p^\mu I_1(m^2)+c_2p^\mu I_2(p^2,m^2,m^2)\ee
The $c$-constants are easy to determine:
\begin{eqnarray}c_1p^2I_1&+&c_2p^2 I_2=\int\frac{d^Dl}{(2\pi)^D}\frac{lp}{[(l+p)^2-m^2](l^2-m^2)}\nonumber\\
&=&\frac{1}{2}\int\frac{d^Dl}{(2\pi)^D}\frac{[(l+p)^2-m^2]-[l^2-m^2]-p^2}{[(l+p)^2-m^2](l^2-m^2)}=
-\frac{p^2}{2}I_2\ ,\end{eqnarray}
hence $c_1=0$ and $c_2=-1/2$.\bs
{\bf Exercise 12.1} Check that the above result agrees with the computations utilizing dimensional regularization and Feynman parameter integration.\es
If we knew in advance that all integrals appearing in $\Sigma_H$ can be expressed in terms of tadpoles and bubbles, we could have started from the ansatz
$$\Sigma_H(p^2)=c_{\Sigma 1}I_1+c_{\Sigma 2}I_2\ .$$
Instead of computing the loop diagrams, we could have tried to determine the $c$-coefficients by comparing the discontinuity of the l.h.s.\ computed by using on-shell amplitudes (unitarity cuts) with the known discontinuities of the r.h.s.
 Since $I_1$ has no discontinuity, we would not obtain any information about $c_{\Sigma 1}$ but
 \be {\rm Disc}_{p^2}\Sigma_H(p^2)=y^2(2p^2-8m^2)\left(\frac{1}{8\pi}\right)\sqrt{1-{4m^2\over p^2}}
 =c_{\Sigma 2}\left(\frac{1}{8\pi}\right)\sqrt{1-{4m^2\over p^2}}, \ee
 hence
$c_{\Sigma 2}=y^2(2p^2-8m^2)$. We conclude that it is possible determine $\Sigma_H(p^2)$ by using the unitarity cut method, up to the tadpole term. Actually, in some theories we know in advance that the tadpole term is absent. In such cases, this on-shell method allows complete determination of $\Sigma_H(p^2)$. As an example consider a supersymmetric theory with the superpotential
\be W=\frac{1}{2}(mX^2+yHX^2+MH^2)\label{spot}\ee
where $H$ and $X$ are supersymmetric chiral multiplets, with $X=(\Phi,\Psi)$. In this theory, the scalar potential
$$V=\left|\frac{\partial W}{\partial X}\right|^2+\left|\frac{\partial W}{\partial H}\right|^2=y^2H^*H\Phi^*\Phi+\dots$$
contains the quartic scalar coupling written on the r.h.s., which gives rise to tadpole contribution that cancels the tadpole part of the fermionic loop. This must be the case because it is known that supersymmetric theories  are free from quadratic ultra-violet divergences.\bs
{\bf Exercise 12.2} Use the on-shell method to express $\Sigma_H$ in terms of the bubble integral $I_2$, in the supersymmetric theory defined by the superpotential $W$ of Eq.(\ref{spot}). You should find that the mass is not renormalized at the one loop level, in agreement with the well-known non-renormalization theorem for the superpotential terms. Caution: A supersymmetric chiral multiplet contains a single chiral fermion, so it contains one-half  degrees of freedom of a Dirac fermion.\es

The treatment of self-energy corrections displays the essence of the unitarity cut method. First, the amplitude is decomposed into a ``basis'' of simple scalar integrals. Then the coefficients are determined by comparing the on-shell discontinuity with the discontinuities of simple integrals. In many cases, in particular in gauge theories, this allows complete determination of the amplitude. In such cases, the amplitudes are called ``cut constructible''.

The starting point is the ``Passarino-Veltman decomposition'' which states that in a renormalizable theory, any one-loop integral, with an arbitrary number of external legs, can be expressed as a linear combination of tadpoles, bubbles, triangles and boxes. The latter are defined as
\begin{eqnarray}
&& I_3(p_1^2,p_2^2,p_3^2; m_1^2,m_2^2,m_3^2)=\int\frac{d^Dl}{(2\pi)^D}\frac{1}{d_1d_2d_3}\\
 && I_4(p_1^2,p_2^2,p_3^2,p_4^2;s_{12},s_{23}; m_1^2,m_2^2,m_3^2,m_4^2)=\int\frac{d^Dl}{(2\pi)^D}\frac{1}{d_1d_2d_3d_4}\ ,
\end{eqnarray}
where
$$d_i=(l+q_i)^2-m_i^2+i0\ ,\quad s_{ij}=(p_i+p_j)^2\ ,\quad q_n=\sum_{i=1}^np_i\ ,\quad q_N=0\ .$$
The triangle and box integrals are ultra-violet finite. They are known in analytic form and at the ${\cal O}(\epsilon^0)$ order they can be expressed in terms of logarithmic and dilogarithmic functions.
The decomposition reads:
\be I_N=\sum_{j_4}c_{4,j_4}I_4^{(j_4)}+\sum_{j_3}c_{3,j_3}I_3^{(j_3)}+\sum_{j_2}c_{2,j_2}I_2^{(j_2)}+
\sum_{j_1}c_{1,j_1}I_1^{(j_1)}+{\cal R}+{\cal O}(\epsilon)\ ,\ee
where the $j$-indices refer to the various distributions of the momenta $p_i$ of the $N$ legs of $I_N$. In the above expression, $\cal R$ denotes possible terms that are  rational functions of the external variables. They have no discontinuities.
\subsection{Example: One-loop correction to four-gluon amplitude}
As an example, we will  compute the one-loop corrections to the four-gluon partial amplitude $A^{(1)}(1^-,2^-,3^+,4^+)$:
\be A^{(1)}(1^-,2^-,3^+,4^+)=c_4I_4+\sum_{j_3}c_{3,j_3}I_3^{(j_3)}+\sum_{j_2}c_{2,j_2}I_2^{(j_2)}
+{\cal R}\ .\label{adis}\ee
In this case, similar to the supersymmetric model discussed before, the tadpole contributions are absent.
We will determine the $c$-coefficients by using the unitarity cut method.

We have two two-particle channels, $s=s_{12}$ (for $s>0$) and $t=s_{23}$ (for $t>0$), available for the cuts. Before analyzing these cuts, let us discuss the relation between the group factors of the one-loop amplitude and of the  tree amplitudes appearing on the two sides of the cut, say in the $s$-channel,
so that (1,2) and (3,4) are located on the opposite sides of the cut. Let us specify to $U(N)$. If we want 1 adjacent to 2 and 3 adjacent to 4, we should consider at least two possibilities:
\begin{eqnarray} \sum_{b,c}2^{\frac{4}{2}}{\rm Tr}(T^{a_1}T^{a_2}T^{b}T^{c})2^{\frac{4}{2}}{\rm Tr}(T^{b}T^{c}T^{a_3}T^{a_4})&=&2^{\frac{4}{2}}{\rm Tr}(T^{a_1}T^{a_2}){\rm Tr}(T^{a_3}T^{a_4})\\
\sum_{b,c}2^{\frac{4}{2}}{\rm Tr}(T^{a_1}T^{a_2}T^{b}T^{c})2^{\frac{4}{2}}{\rm Tr}(T^{c}T^{b}T^{a_3}T^{a_4})&=&2^{\frac{4}{2}}N\,{\rm Tr}(T^{a_1}T^{a_2}T^{a_3}T^{a_4})\label{sing}\end{eqnarray}
We see that not only single-trace, but also double trace factors appear at the one loop level. Since we are interested in the partial amplitude associated to the single-trace ${\rm Tr}(T^{a_1}T^{a_2}T^{a_3}T^{a_4})$, see Eq.(\ref{sing}), the relevant contribution is
\begin{eqnarray} &&{\rm Disc}_{s} A^{(1)}(1^-,2^-,3^+,4^+)\label{sdis2}\\ &&=-\sum_{h_b,h_c}\int d\mu\, 
A(1^-,2^-,(-b)^{-h_b},(-c)^{-h_c})\times A(c^{h_c},b^{h_b},3^+,4^+)\ ,\nonumber
\end{eqnarray}
where $d\mu=dLIPS_2(l_b+l_c=p_1+p_2)$.

We can consider not only pure Yang-Mills theory but also its supersymmetric extensions with fermions (gauginos) and possibly scalars in the adjoint representation. A notable example is ${\cal N}=4$ SYM with four gauginos and six gauge scalars, which has the remarkable property of being ultra-violet finite. In either case, no fermions or scalars, but only gauge bosons with $h_b=h_c=-1$ appear in the $s$-channel discontinuity (\ref{sdis2}). The integrand of Eq.(\ref{sdis2}) is given by
\begin{eqnarray}  A(1^-,2^-,(-b)^{+},(-c)^{+})\times A(c^{-},b^{-},3^+,4^+)&=&\\[1mm]
 \frac{\langle 12\rangle^4}{\langle 12\rangle\langle 2b\rangle\langle bc\rangle\langle c1\rangle}
\frac{\langle cb\rangle^4}{\langle cb\rangle\langle b3\rangle\langle 34\rangle\langle 4c\rangle}
&=&-\underbrace{\frac{\langle 12\rangle^4}{\langle 12\rangle\langle 23\rangle\langle 34\rangle\langle 41\rangle}}_{A(1^-,2^-,3^+,4^+)}
\frac{\langle bc\rangle^2\langle 23\rangle\langle 41\rangle}{\langle 2b\rangle\langle b3\rangle\langle 4c\rangle\langle c1\rangle},\nonumber
\end{eqnarray}
where we factored out the tree amplitude.\bs
{\bf Exercise 12.3} Show that
\be -\frac{\langle bc\rangle^2\langle 23\rangle\langle 41\rangle}{\langle 2b\rangle\langle b3\rangle\langle 4c\rangle\langle c1\rangle}=\frac{st}{(p_1-l_c)^2(p_3+l_b)^2}\label{idd}\ee\es
In order to compare Eq.(\ref{sdis}) with the discontinuity of the r.h.s.\ of (\ref{adis}), we need to compute ${\rm Disc}_{s}I_4$, ${\rm Disc}_{s}I_3$ and ${\rm Disc}_{s}I_2$.
\begin{figure}[h!]
\begin{center}
\includegraphics[scale=0.6]{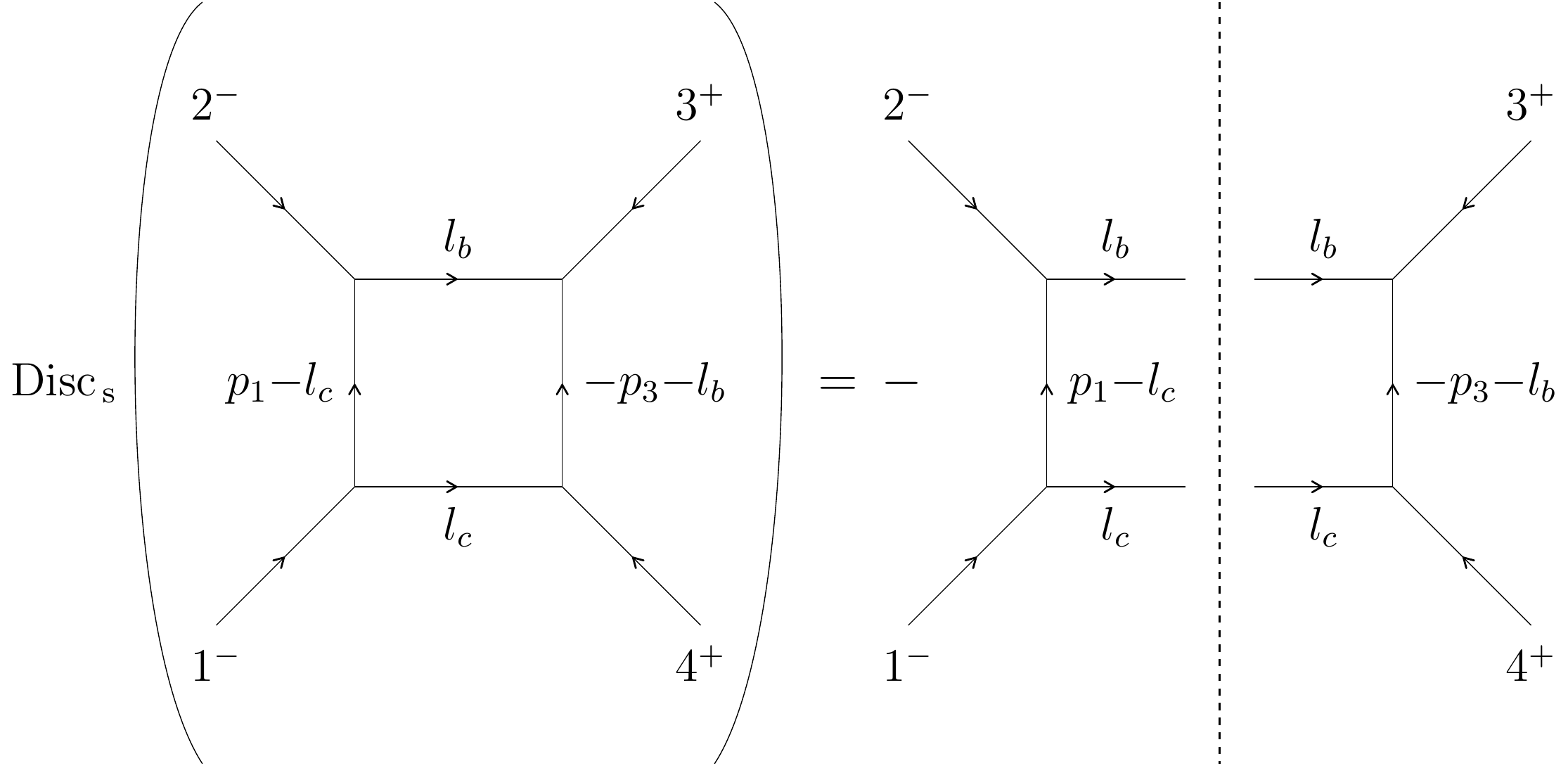}
\end{center}
\caption{$s$-channel cut of the box integral.}
\end{figure}
For the box integral, see Fig.9,
\be {\rm Disc}_{s}I_4=-\int d\mu\frac{1}{(p_1-l_c)^2}\frac{1}{(p_3+l_b)^2}\ ,\ee
After collecting all these results together, we obtain
\be A^{(1)}(1^-,2^-,3^+,4^+)=Nst A(1^-,2^-,3^+,4^+)I_4(s,t)+\makebox{triangles}+\makebox{bubbles}
+{\cal R}\ ,\label{bdis}\ee
with the r.h.s.\ involving only those triangles and bubbles that have no $s$-channel discontinuities.

In order to determine the remaining coefficients, we move to the $t$-channel cuts. Now the relevant discontinuity is shown in Fig.10:
\begin{eqnarray} &&{\rm Disc}_{t} A^{(1)}(1^-,2^-,3^+,4^+)\\ &&=-\sum_{h_b,h_c}\int d\mu\, A(2^-,3^+,(-c)^{-h_c},(-b)^{-h_b})\times A(b^{h_b},c^{h_c},4^+,1^-)\  ,\nonumber\end{eqnarray}
where $d\mu=dLIPS_2(l_b+l_c=p_2+p_3)$.
\begin{figure}[h!]
\begin{center}
\includegraphics[scale=0.6]{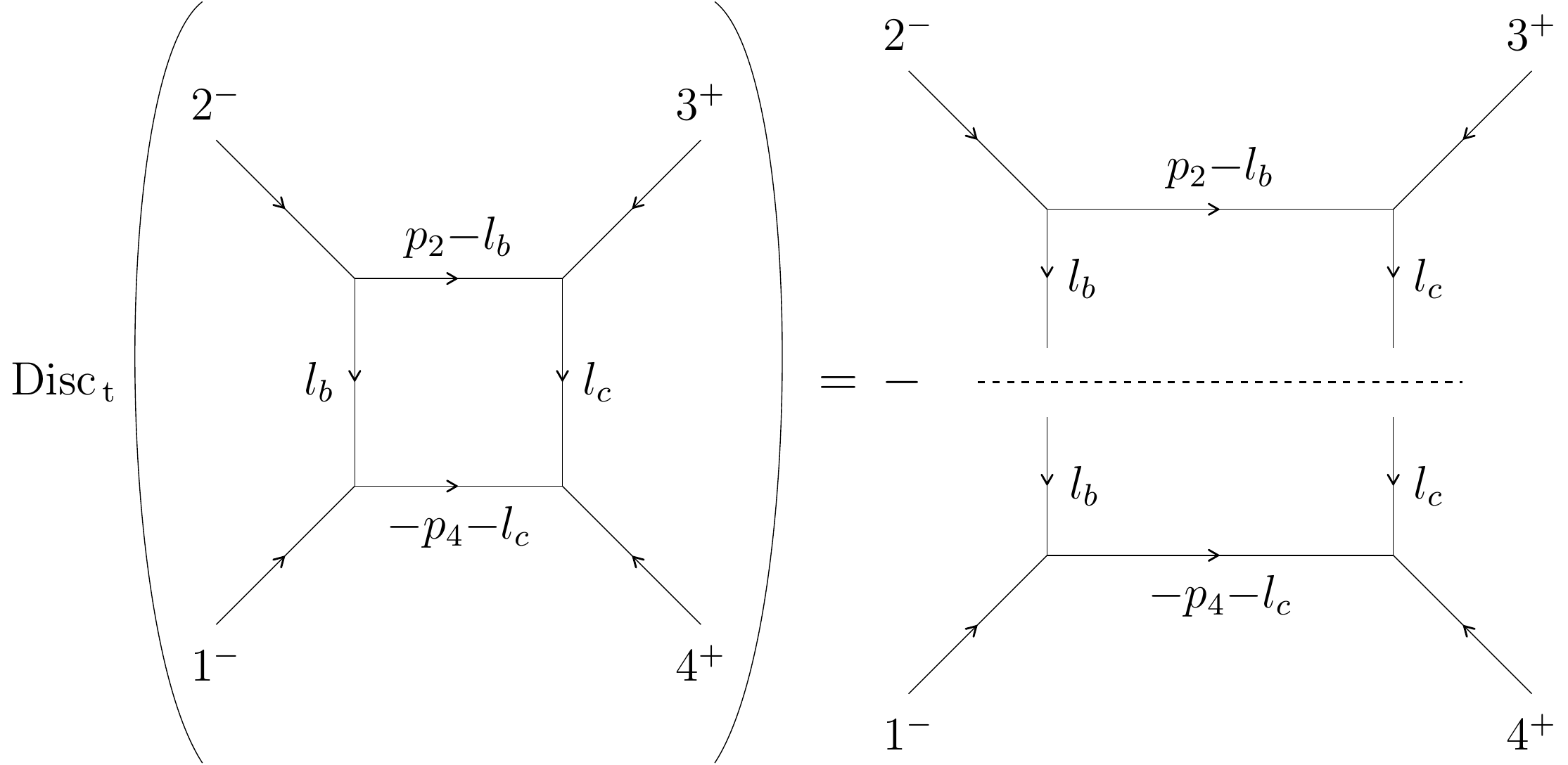}
\end{center}
\caption{$t$-channel cut of the box integral.}
\end{figure}
In this case, there are more helicity configurations possible:
\begin{eqnarray} &&{\rm Disc}_{t} A^{(1)}(1^-,2^-,3^+,4^+)\label{tdis}\\ &&= ~-\!\!\!\!\!\!\!\!\sum_{\scriptscriptstyle h=\{-1,-1/2,0,1/2,1\}}\int d\mu\, A(2^-,3^+,(-c)^{h},(-b)^{-h})\times A(b^{h},c^{-h},4^+,1^-)\ ,\nonumber\end{eqnarray}\bs
{\bf Exercise 12.4} Use Eq.(\ref{ddh1}) to show that at the tree level:
\begin{eqnarray}
A(2^-,3^+,(-c)^{h},(-b)^{-h})&=&D(-h)\frac{\langle b2\rangle^{2+2h}\langle c2\rangle^{2-2h}}{\langle 23\rangle\langle 3c\rangle\langle cb\rangle\langle b2\rangle}\label{aup}\\[2mm]
A(b^{h},c^{-h},4^+,1^-) &=&D(h)\frac{\langle c1\rangle^{2+2h}\langle b1\rangle^{2-2h}}{\langle bc\rangle\langle c4\rangle\langle 41\rangle\langle 1b\rangle}\ ,\label{aup1}
\end{eqnarray}
where $D(h)$ is written in Eq.(\ref{ddh}). Note that $D(h)D(-h)=(-1)^{2h}$.\es

Consider a gauge theory with $n_f$ gauginos and $n_s$ (real) gauge scalars in the adjoint representation.
Then in the integrand on the r.h.s.\ of  (\ref{tdis}), the products of numerators of Eqs.(\ref{aup}), (\ref{aup1}) combine to:
\begin{eqnarray}
&& \langle b2\rangle^{4}\langle c1\rangle^{4}+\langle c2\rangle^{4}\langle b1\rangle^{4}\nonumber \\[2mm]
&& -~n_f\Big(\langle b2\rangle^{3}\langle c1\rangle^{3}\langle c2\rangle\langle b1\rangle+\langle b2\rangle\langle c1\rangle\langle c2\rangle^3\langle b1\rangle^3\Big)\label{tdisss}\\[2mm] &&
+~n_s\langle b2\rangle^{2}\langle c1\rangle^{2}\langle c2\rangle^2\langle b1\rangle^2\nonumber
\end{eqnarray}
The case of $n_f=4$, $n_s=6$, as in ${\cal N}=4$ SYM, is particularly interesting because the r.h.s.\ combines
then to the quartic
\be
\Big(\langle b2\rangle\langle c1\rangle-\langle c2\rangle\langle b1\rangle\Big)^4=\langle 12\rangle^{4}\langle bc\rangle^{4}
\ee
so that
\begin{eqnarray} {\rm Disc}_{t} A^{(1)}(1^-,2^-,3^+,4^+)\label{tdist}&=& \int d\mu\, A(1^-,2^-,3^+,4^+)
\frac{\langle bc\rangle^2\langle 12\rangle\langle 34\rangle}{\langle 3c\rangle\langle c4\rangle\langle 1b\rangle\langle b2\rangle}\label{dta}\\[2mm]
&=& -\int d\mu\, A(1^-,2^-,3^+,4^+)\frac{st}{(p_2-l_b)^2(p_4+l_c)^2}\nonumber
\end{eqnarray}
On the other hand
\be {\rm Disc}_{t}I_4=-\int d\mu\frac{1}{(p_2-l_b)^2}\frac{1}{(p_4+l_c)^2}\ .\label{dti}\ee
After comparing Eqs.(\ref{bdis}), (\ref{dta}) and (\ref{dti}), we confirm the coefficient of the box integral in
Eq.(\ref{bdis}) and conclude that in ${\cal N}=4$ SYM
\be A^{(1)}(1^-,2^-,3^+,4^+)=Nst A(1^-,2^-,3^+,4^+)I_4(s,t)
~+~{\cal R}\qquad ({\cal N}=4 ~{\rm SYM})\ ,\label{bbdis}\ee
i.e.\ that the triangle and bubble contributions are absent. In addition, it can be shown that ${\cal R}=0$ in
${\cal N}=4$ SYM. It should be also noted that
 in a generic gauge theory with arbitrary numbers of fermions and scalars, the discontinuity computed by using Eqs.(\ref{tdis}) and (\ref{tdisss}) can be matched with Eq.(\ref{bdis}) by including the bubbles only; there are no triangle contributions.
 \renewcommand{\refname}{\large Recommended Reading for Section 12}

\end{document}